\documentstyle[prb,twocolumn,aps]{revtex}
\begin{document}
\draft
\title{Superconductivity in the Hubbard model with correlated hopping:
Slave-boson study}
\author { Bogdan R. Bu{\l}ka\cite{ebb}}
\address{Institute of Molecular Physics,
Polish Academy of Sciences,   \\
ul. Smoluchowskiego 17, 60-179 Pozna\'n, Poland}
\date{Received \hspace{5mm} }
\date{\today}
\maketitle
\begin{abstract}
The slave boson mean-field studies of the ground state of the Hubbard model
with correlated hopping were performed. The approach qualitatively recovers
the exact results for the case of the hopping integral $t$ equal to the
correlated hopping integral $X$. The phase diagram for the strongly correlated
state with only singly occupied sites, the weakly correlated state, where
single and double occupation is allowed, and for the superconducting state,
was determined for any values of $X$ and any electron concentration $n$. At
the half-filled band ($n=1$) a direct transition from the superconductor to the
Mott insulator was found. In the region of strong correlations the
superconducting solution is stable for $n$ close to 1, in contrast to the case
of weak correlations, in which superconductivity occurs at $n\approx 0$ and
$n\approx 2$. We found also that strong correlations change characteristics of
the superconducting phase, e.g. the gap in the excitation spectrum has a
nonexponential dependence close to the point of the phase transition.
\end{abstract}

\section{INTRODUCTION}

      	Over 10 years passed from the discovery in cuprates superconductivity
with a high critical temperature, however, its nature and the mechanism
leading to superconductivity have not been yet identified in a conclusive
manner. There are many experimental facts, which suggest a significant role of
magnetic and charge-charge interactions in formation of the superconducting
phase (e.g. in cuprates and bismuthates, respectively). The simplest model
describing the physical situation in such compounds and taking into account
electronic correlations on the lattice is the Hubbard model. An important
question, which arises is a mechanism of formation of Cooper pair in the
presence of strong Coulomb onsite repulsion. In some circumstances even not
strong intersite Coulomb interactions, as that one for correlated hopping,
overwhelm onsite repulsion and lead to superconductivity for hole
hoping.~[\cite{MRR,Kiv,Bae,Hir89,M89,App}] Recently the 1D Hubbard model with
correlated hopping was analyzed for the case when the hopping integral $t$ is
equal to the correlated hopping integral
$X$.~[\cite{Strack,Ovch,Elsser,Aligia94,Aligia,Arrachea,Kollar,OGSA}] Exact
results were obtained by using the conservation of the number of doubly
occupied sites in the model and mapping it onto a system of spinless free
fermions with twisted boundary conditions~[\cite{Elsser}], or in  a equivalent
way, by expression of the model by slave bosons, construction of the eigenstates
and minimizing the energy with respect of number of doubly occupied
sites.~[\cite{Aligia94}] The superconducting phase was found as a ground state
for the onsite Coulomb integral $U<-4t\cos(\pi n)$ ($n$ is the concentration
of electrons). In the case of strong attraction, $U<-4t$, in the ground state
the sites are occupied by two electrons or are empty, and there are no singly
occupied sites. In the region $U>-4t\cos(\pi n)$ there are only singly occupied
sites allowed, what corresponds the situation $U=\infty$ in the pure Hubbard
model ($X=0$).

	The other type of intersite interactions may destabilize
superconductivity. The density-density coulomb interactions lead to the charge
density wave (CDW) ordering for $n=1$. The stability conditions of the CDW
were studied in an exact manner for $n=1$ and $X=t$ by Strack et
al.~[\cite{Strack}] and Ovchinnikov~[\cite{Ovch}]. The stability of saturated
ferromagnetism in the presence of the exchange Heisenberg and the pair hopping
terms was studied by Kollar et al.~[\cite{Kollar}] using the generalized
Nagaoka's theorem for $n=1$ with one hole. The most general situation was
analyzed by de Boer and Schadschneider~[\cite{OGSA}]. Using the optimal ground
state approach they determined~[\cite{OGSA}] sufficient conditions for the uniform
and modulated superconducting state, CDW, AF and ferromagnetic state for the
extended Hubbard model, which included all type of nearest neighbor (n.n.)
interactions on the arbitrary lattice at the case $X=t$.

	Deviation from the special point $X=t$ removes the degeneracy of the
ground state. There are not known exact analytical methods, which work apart
from this point for arbitrary values of $X$, and may be use to determine the
ground state. The Hartree-Fock approximation (HFA) may be useless for such
studies as it neglects correlations of electrons and can not describe the Mott
transition at $n=1$ as well as the state with singly occupied sites (the
strongly correlated state) for $n\ne 1$. The slave boson mean-field
approximation~[\cite{Kotliar}] (SBMFA) may be use for this purpose as it is
equivalent to the Gutzwiller approximation to the Gutzwiller wave
function.~[\cite{Gutzwiller}] The approach has been extensively used for the
Hubbard model and its various
generalizations.~[\cite{Has,H89,La,Den,Dzierzawa,Evans,Av,BB,Feshke,Sofo,rob}] 
For the repulsive case $U>0$ the stability of different magnetic orderings:
ferromagnetic, ferrimagnetic, antiferromagnetic and spiral structures, as well as the
CDW state were investigated.~[\cite{Dzierzawa,Evans,Av,BB,Feshke}] An influence
of a static distortion of the lattice on the stability of the magnetic phase
was also studied.~[\cite{Feshke}] Properties of the superconducting state
were studied either by a transformation of the attractive Hubbard model to its
repulsive analog and solving the stability conditions for the
antiferromagnetic phase in the presence of an external magnetic
field~[\cite{Sofo}], or by means of the spin- and the charge-rotationally
invariant representation of slave bosons~[\cite{BB,rob}]. The SBMFA
results~[\cite{rob}] for the ground state (the free energy, the gap in the
excitation spectrum, the critical magnetic field and the coherence length)
were compared with those obtained in the HFA and by the self-consistent
second order perturbation theory in the weak coupling regime
($U/t\ll 1$)~[\cite{VD,Fr}] as well as with those obtained using perturbations
approaches in the strong coupling regimes ($U/t\gg 1$)~[\cite{UU}]. The SBMFA,
in contrast to the HFA, gives~[\cite{rob}] credible results in the whole
interaction range and for any electron concentration. The one orbital Hubbard
model with a generalized correlated hopping term has been derived as an
effective model from the three orbital model for CuO$_2$ plane of cuprates by
Simon and Aligia~[\cite{Simon93}]. They also used the SBMFA, but confined
themselves to the analyze of the Mott transition for some restricted range of
parameters of the model.

	The purpose of the present work is to determine the ground state of
the Hubbard model with correlated hopping and to study its basic
characteristics in  for whole range of parameters $X$ and $U$, as well as
for any electron
concentration $n$. We are convinced that the SBMFA gives reliable qualitative
description of the physical situation, especially the strongly correlated
normal phase (with only singly occupied sites) and the superconducting phase
in the presence of correlations. After the introduction of the model and the
method in the Section 2, we compare, in the Section 3a, the SBMFA results for
$X=t$ with those obtained in the HFA as well as with rigorous solutions.
It will be shown
that the SBMFA results are very close to the exact ones, and differ from the
HFA. The case of the half-filled electronic band ($n=1$) is considered
separately in the Section 3b. We will determine the phase diagram for the
superconducting state in the space of the parameters of the model. Apart from
the phase transition from the superconducting to the metallic state for small
values of $X$, we expect to find the phase transition from the superconducting
state directly to the Mott insulator in a finite range of $X$ close to the
point $X=t$. We will undertake investigations of characteristics of the
superconductor-to-Mott insulator transition, as due to strong correlation
effects they may be different from those known for the superconductor-metal
transition. The Section 3c is devoted to the analyze of the stability of the
solutions for the superconducting and the normal phase in the most general
case $n\ne 1$ and $X\ne t$. We predict the strongly correlated metallic phase
around the point $X=t$. The region of the stability for this phase depends
on the concentration of electrons $n$, and it can be different for doping by holes
and electrons. The phase transition to the superconducting state will be
studied in the whole range of $X$ and $n$. We will compare the SBMFA results
with those obtained by the HFA. The differences may be quite large, especially close to the point $X=t$, where the HFA
does not work properly and can not reproduce the exact
results.~[\cite{Strack,Ovch,Elsser,Aligia94,Aligia,Arrachea,Kollar,OGSA}]
Discussion and comparison with other approaches will be performed the last
Section.

\section{SLAVE-BOSON REPRESENTATION OF THE HUBBARD MODEL WITH CORRELATED
HOPPING}

	The Hamiltonian of the Hubbard model with correlated hopping is
given by
\begin{eqnarray}\label{1}
H =
&&\sum _{<i,j>,\sigma}(c_{i\sigma}^{\dag}c_{j\sigma}
+c_{j\sigma}^{\dag}c_{i\sigma}
)[-t+X(n_{i-\sigma}+n_{j-\sigma})]\nonumber\\
&&+U\sum_in_{i\uparrow}n_{i\downarrow} ,
\end{eqnarray}
where the sum is restricted to n.n. sites. The parameters of the Hamiltonian
(\ref{1}) were determined by Hubbard~[\cite{H}] as $U=20$ eV and $X=0.5$ eV from
purely Coulombic origin. Screening effects may considerably reduced their
values,~[\cite{H}] especially in highly anizotropic systems.~[\cite{GC}] An
estimation for cuprates and bismuthates gives $U=3\div 5$ eV, $X=0.1\div 0.3$
eV and $t=0.1\div 0.2$ eV. There are several mechanisms, which may lead to
a further reduction of $U$ and even to overcome it. One of them is a polaronic
mechanism, which can be realized in a narrow-band electronic system coupled
to high-frequency intramolecular vibrations or cation-ligand
vibrations.~[\cite{MRR}] The short-range attraction can result from a coupling
between electrons and quasi-bosonic excitations such as excitons or
plasmons.~[\cite{ex}] Another possibility is a purely electronic origin resulting
from a coupling of electrons in a many orbital model as has been derived
by Simon and Aligia.~[\cite{Simon93}]. Therefore, the Hamiltonian (\ref{1})
may be treated as an effective one, where its parameters $U$ and $X$ may
change in the whole range, they may be also negative. The hopping integral
$t$ is taken positive (without losing generality of the considerations).

	We want to express the Hamiltonian (\ref{1}) in the slave-boson
representation, which is the spin- and the charge-rotationally
invariant.~[\cite{BB,FW}] The singly occupied states $|\sigma\rangle_i$ are
expressed by the Bose operators $p_{i,\sigma\sigma^\prime}^{\dag}$ and the Fermi
operators $f_{i,\sigma}^{\dag}$ as
\begin{equation}\label{2}
|\sigma\rangle_i = \sum_{\sigma^\prime}p_{i,\sigma\sigma^\prime}^{\dag}
f_{i,\sigma^\prime}^{\dag} |vac\rangle\;\;\:  \mbox{($\sigma = \uparrow,
\downarrow$)}.
\end{equation}
The doubly occupied state $|2\rangle_i$ and the state $|0\rangle_i$,
corresponding to an empty site, form a doublet $|\rho\rangle_i$ ($\rho$ =
2,0), which may be expressed by the Bose operators $b_{i,\rho\rho^\prime}^{\dag}$
as
\begin{equation}\label{3}
|\rho\rangle_i = \sum_{\rho^\prime}b_{i,\rho\rho^\prime}^{\dag}
\psi_{i,\rho^\prime}^{\dag} |vac\rangle,
\end{equation}
where
\begin{equation}\label{4}
\psi_{i,\rho}^{\dag} = \pmatrix{f_{i,\uparrow}^{\dag} f_{i,\downarrow}^{\dag}\cr1}.
\end{equation}
In order to operate in the physical part of the extended Hilbert space we
introduce the following constraint:
\begin{equation} \label{5}
2\sum_{\rho\rho\prime}b_{i,\rho\rho^\prime}^{\dag}b_{i,\rho^\prime\rho}+2\sum_{
\sigma\sigma^\prime}p_{i,\sigma\sigma^\prime}^{\dag}p_{i,\sigma^\prime\sigma}=1.
\end{equation}
It ensures that each site is occupied by exactly one slave-boson. The Hubbard
operators are given by
\begin{eqnarray}\label{6}
&& X_i^{\rho_1\rho_2}\equiv |\rho_1\rangle_{\!i\,\,i}\!\langle\rho_2| =
2\sum_{\rho} b_{i,\rho_1\rho}^{\dag} b_{i,\rho\rho_2}\;,\nonumber\\
&& X_i^{\sigma_1\sigma_2}\equiv |\sigma_1\rangle_{\!i\,\,i}\!\langle\sigma_2| =
2\sum_{\sigma} p_{i,\sigma_1\sigma}^{\dag} p_{i,\sigma\sigma_2}\;,\nonumber\\
&& X_i^{\sigma\rho}\equiv |\sigma\rangle_{\!i\,\,i}\!\langle\rho| =
\sum_{\sigma'} p_{i,\sigma\sigma'}^{\dag}f_{i,\sigma'}^{\dag} \sum_{\rho'}
b_{i,\rho'\rho}
\psi_{i,\rho'}\nonumber\\
&&\quad\quad = \sum_{\sigma'} p_{i,\sigma\sigma'}^{\dag}(b_{i,0\rho}
f_{i,\sigma'}^{\dag}-\sigma'b_{i,2\rho}f_{i,-\sigma'}) .
\end{eqnarray}
By means of these operators any extended Hubbard Hamiltonian may be
expressed, especially the Hamiltonian (1), which takes the form
\begin{eqnarray}\label{7}
H = &&\sum_{<i,j>,\sigma}[t_1 X_i^{\sigma 0}X_j^{o\sigma}+ t_2 X_i^{2\sigma}
X_j^{\sigma 2}\nonumber\\
&&\quad\quad+\sigma t_{12} (X_i^{2-\sigma}X_j^{0\sigma}+X_i^{\sigma 0}
X_j^{-\sigma 2}) +h.c.]\nonumber\\
&& + U \sum_{i} X_i^{22} ,
\end{eqnarray}
where $t_1 = -t$, $t_2= -t+2X$, $t_{12} = -t+X$. For arbitrary values of the
parameters $t_1$, $t_2$ and $t_{12}$ it is the most general form of the
Hamiltonian with correlated hopping. If $t_{12}=0$ the Hamiltonian commutes
with the total spin, the number of doubly occupied sites $D=\sum_{i}
n_{i\uparrow}n_{i\downarrow}=\sum_i X_i^{22}$ and the $\eta$ operators:
\begin{eqnarray}\label{7a}
&&\eta= \sum_i c_{i\uparrow}c_{i\downarrow}=-\sum_iX_i^{02},\nonumber\\
&&\eta^{\dag}=\sum_i
c_{i\downarrow}^{\dag}c_{i\uparrow}^{\dag}=-\sum_i X_i^{20},\nonumber\\ &&\eta^z=\frac{1}{2}
\sum_i(n_{i\uparrow}+n_{i\downarrow}-1)\;.
\end{eqnarray}
This fact was used in finding the ground state in the exact
manner.~[\cite{Elsser,Aligia94}]

	We want to study the Hamiltonian (\ref{7}) in the mean field
approximation, but instead simply to exchange the Bose operators in
Eqs.(\ref{6}) by the c-numbers we renormalize properly the operators $X$. We
use the fact that representation of any operator by the slave operators is not
uniquely defined, but only up to operator factor whose eigenvalue is unity in
the physical space. The renormalized operators $X$ may be expressed as
\begin{equation}\label{8}
X_i^{\sigma\rho} = A_i^{0\rho}f_{i\sigma}^{\dag}-\sigma A_i^{2\rho}f_{i-\sigma} ,
\end{equation}
where the $2\times2$ matrix $\underline{A}$ is
\begin{equation}\label{9}
\underline{A} = \underline{p}^{\dag}\underline{L}\underline{R}\underline{b} ,
\end{equation}
and
\begin{eqnarray}\label{10}
&& \underline{L} = [1-\underline{p}^{\dag}\underline{p}-\underline{b}^{\dag}
\underline{b}]^{-1/2}\;,\;
\underline{R} = [1-\underline{\tilde{p}}^{\dag}\underline{\tilde{p}}-
\underline{\tilde b}^{\dag}\underline{\tilde b}]^{-1/2}\;,\nonumber\\
&& \underline{b} = \pmatrix{b_{22}&b_{20}\cr b_{02}&b_{00}}\;,\;
\underline{\tilde b} = \pmatrix{b_{00}& -b_{20} \cr -b_{02}&b_{22}}\;,\nonumber\\
&&\underline{p} = \pmatrix{p_{\uparrow\uparrow}&p_{\uparrow\downarrow}\cr
p_{\downarrow\uparrow}&p_{\downarrow\downarrow}}\;,\;
\underline{\tilde p} = \pmatrix{p_{\downarrow\downarrow}& -p_{\uparrow
\downarrow} \cr -p_{\downarrow\uparrow}&p_{\uparrow\uparrow}}\;.
\end{eqnarray}

	Now, we may perform the mean-field studies of the stability of the
superconducting phase. We determine the lowest energy solution from the
minimum of the free energy $F$
\begin{eqnarray}\label{11}
F=&& F_f+F_b =\nonumber\\
&& -\frac{1}{\beta}\sum_{\bf {k}} [\ln\lbrace 1+\exp(\beta E_k)
\rbrace+\ln\lbrace1+\exp(-\beta E_k)\rbrace] \nonumber\\
&& +\frac{UN}{2}(b^2+2\delta)-\lambda_0(1+2\delta)N-2\lambda_{\Delta}\Delta N ,
\end{eqnarray}
where $F_f$ and $F_b$ are the fermionic and bosonic parts of the free energy,
$\Delta$ is the superconducting order parameter defined by
\begin{equation}\label{12}
\Delta \equiv \langle c_{i\downarrow}c_{i\uparrow}\rangle = 2\sum_{\rho}
\langle b_{i\rho 2}^{\dag}b_{i 0\rho}\rangle ,
\end{equation}
and $E_k = \sqrt{(-2\tilde t\eta_k+\lambda_0)^2+(2r\eta_k+ \lambda_{
\Delta})^2}$ is the dispersion curve of fermions described by the Hamiltonian
(\ref{7}) in the SBMFA, i.e. by
\begin{eqnarray}\label{13}
H = \sum_k [&&(-2\tilde t\eta_k+\lambda_0)\sum_{\sigma}f_{k\sigma}^{\dag}f_{k\sigma}\nonumber\\
&&+ (2r\eta_k+\lambda_{\Delta})(f_{k\uparrow}^{\dag}f_{k\downarrow}^{\dag}+ h.c.)] .
\end{eqnarray}
The following notation is used
\begin{eqnarray}\label{14}
\tilde t =&& -a^2[b_0^4(t_1+t_2+2t_{12})+2\delta b_0^2(-t_1+t_2)\nonumber\\
&&+(\delta^2-\Delta^2) (t_1+t_2-2t_{12})],\nonumber\\
r =&& a^2[b_0^2(-t_1+t_2)+\delta(t_1+t_2-2t_{12})]2\Delta ,
\end{eqnarray}
where
\begin{eqnarray}
a^2 =&& p^2/[b_0^2(1-4\delta^2-4\Delta^2)]\;,\nonumber\\
b_0^2=&& (b^2+\sqrt{b^4-4\delta^2-
4\Delta})/2\;,\nonumber\\
b^2 =&& 2\langle b_{i22}^{\dag}b_{i22}+b_{i00}^{\dag}b_{i00}+b_{i20}^{\dag}
b_{i02}+b_{i02}^{\dag}b_{i20}\rangle\;,\nonumber\\
p^2 = &&2\langle p_{i\uparrow\uparrow}^{\dag}p_{i\uparrow\uparrow}+p_{i\downarrow
\downarrow}^{\dag}p_{i\downarrow\downarrow}\rangle\;,\nonumber
\end{eqnarray}
the electron concentration is 
\begin{equation}
n=1+2\delta=1+2(\langle b_{i22}^{\dag}b_{i22}\rangle - \langle b_{i00}^{\dag}b_{i00}\rangle)\;,\nonumber
\end{equation}
$\eta_k = \sum_{\alpha}\cos k_{\alpha}$, $N$ is the number of the
lattice sites. The Lagrange multipliers $ \lambda_0$ and $\lambda_{\Delta}$
are for the constraints, which ensure the number of electrons and the
superconducting order parameter $\Delta$.

	The free energy (\ref{11}) is the function of the variational
parameters: $b^2$, $\Delta$, $\lambda_0$ and $\lambda_{\Delta}$. The
parameter $b^2$ changes in the range from $2\sqrt{\delta^2+\Delta^2}$ to 1,
and the superconducting order parameter $\Delta \le \frac{1}{2}\sqrt{1-4
\delta^2}=\frac{1}{2}\sqrt{n(2-n)}$. In the normal phase ($\Delta=0$) $b^2$
describes the concentration of doubly occupied sites, which is 0 if $b^2=2|
\delta|$, and maximal equal to $n/2$ for $b^2=1$ (according to notation used
by Kotliar and Ruckenstein~[\cite{Kotliar}] in the normal phase  $b^2=d^2+e^2=
2d^2+1-n$, where $d^2$ and $e^2$ are the concentration of doubly occupied
sites and empty sites, respectively). At $b^2=2|\delta|$ there are no doubly
occupied sites, the state is paramagnetic metal with a maximal value of the
local magnetic moment. Such a state corresponds the situation for $U=\infty$
in the pure Hubbard model (the case $X=0$) and we call it the strongly
correlated normal state (SC). For $2|\delta|<b^2<1$ there are allowed doubly
as well as singly occupied sites. We call it the weakly correlated normal
state (WC) in order to differentiate it from the SC state. The local magnetic
moment decreases with increasing $b^2$, and disappears at all for $b^2=1$, i.e.
when only doubly occupied sites are allowed. For the pure Hubbard model ($X$=0)
it happens in the SBMFA~[\cite{rob}] for strong attraction ($U<0$), when the
effective bandwidth becomes zero, the system is then a diamagnetic insulator.
Comparison of the SBMFA free energy with that one obtained by the perturbation
expansion in the limit $|U|\gg t$ shows~[\cite{rob}] that there is still a part
corresponding to mobile particles in a band, what neglects the SBMFA. For
large $|U|/t$ the Cooper pairs have very small size, which is restricted to a
single lattice site, and they form a Bose liquid. The system in the normal
phase is a diamagnetic metal, in which electronic transport is due to motion
of these local pairs (hard core bosons with charge 2e). Condensations of the
bosons leads to superconductivity~[\cite{MRR}].

\section{THE SBMFA ground state characteristics}

\subsection{The  $t = X$ Case}

	Here, we want to analyze the ground state solutions in the SBMFA. In
order to see validity of the present approach we perform calculations for
$X=t$ and compare them with the results obtained in the exact way for the 1D
chain~[\cite{Elsser}]. The considered case will be a reference system for
further studies. It is worth to notice that in high temperature
superconductors the parameters are estimated as $X=0.1\div 0.3$ eV and $t=0.1
\div 0.2$ eV, what is close the considered case.

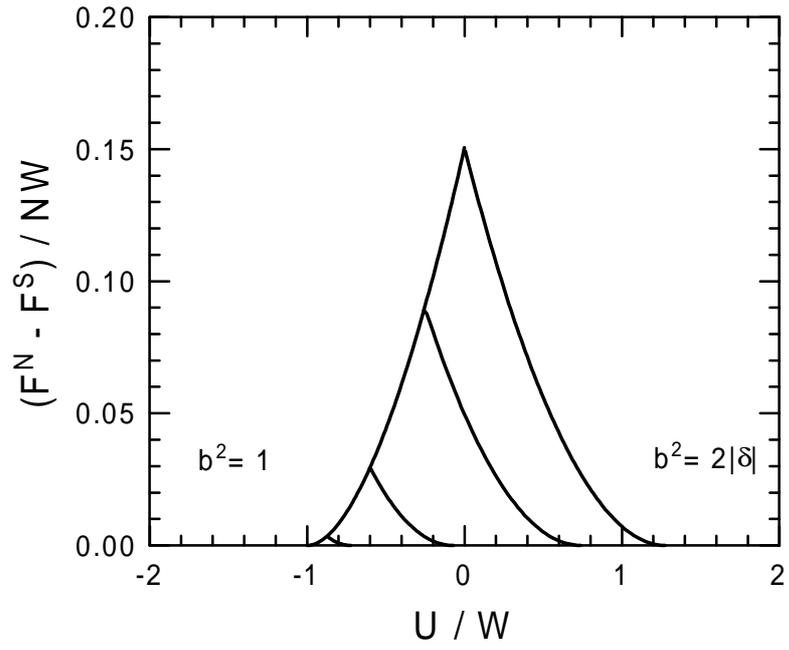
\begin{figure}
\caption{Free energy difference of the normal and the superconducting state
calculated with in the SBMFA at $n=1, 0.8, 0.5, 0.2$  (from top) for the 1D
Hubbard model with correlated hopping at $X=t$. It is shown the value of $b^2$
in the normal phase.}
\label{f1}
\end{figure}

	At $X=t$ one has an electron-hole symmetry, which in general (for t$
\neq X$) is not kept. The parameters of the Hamiltonian (\ref{13}) are now
\begin{equation}\label{15}
\tilde t = -4a^2b_0^2t\delta \;\mbox{ and }\;r=4a^2b_0^2t\Delta .
\end{equation}
To determine the free energy (\ref{11}) we first transform the variables to
\begin{eqnarray}\label{16}
&&\overline\delta = (\tilde t\delta+r\Delta)/\rho\; ,\; \overline\Delta = (r
\delta-\tilde t\Delta)/\rho ,\nonumber\\
&&\overline\lambda_0 = (\tilde t\lambda_0+r\lambda\Delta)/(2\rho^2)\;,\;
\overline \lambda_S = (\lambda_0r-\lambda_{\Delta}\tilde t)/(2\rho^2) ,
\end{eqnarray}
where $\rho^2 = \tilde t^2 + r^2$. It corresponds to a rotation in the space
of the parameters $\delta$ and $\Delta$. The dispersion curve is expressed as
$E_k = 2\rho \sqrt{(\eta_k+ \overline \lambda_0)^2+ \overline \lambda_S^2}$.
For $\tilde t$ and $r$ given by (\ref{15}) the parameter $\overline \Delta
=0$, what implies $\overline \lambda_S = 0$ as well. The gap in the excitation spectrum, defined as $E_g = 2\max(E_k)$, is also equal to zero in this case.
The free energy of the superconducting state for the 1D chain at $T= 0$ is
\begin{equation}\label{17}
F^S/N = -2\rho\frac{1}{\pi}\int_0^{\pi}dk |\eta_k+\overline\lambda_0| +
\frac{U}{2}(b^2+2\delta)-4\overline\lambda_0\overline\delta \rho .
\end{equation}
The minimum of $F$ is at $\overline\lambda_0 = -2\overline\delta$, $b^2 = 2|
\overline\delta|$ and the solution may be expressed in the parametric way as
a function of $\delta$ and $|\overline \delta| = \sqrt{\delta^2+\Delta^2}$
($|\delta|\le |\overline\delta|\le 1/2$)
\begin{eqnarray}\label{18}
F^S/N =&& -\frac{16t|\overline\delta|}{\pi(1+2|\overline\delta|)}\sin\frac{
\pi}{2}(1-2|\overline\delta|) +U (|\overline\delta|+\delta),\nonumber\\
\label{19}\\
U =&& \frac{16t}{1+2|\overline\delta|}\lbrack\frac{\sin\frac{\pi}{2}(1-2|
\overline\delta|)}{\pi(1+2|\overline\delta|)}-|\overline\delta|\cos\frac{\pi}
{2}(1-2|\overline\delta|)\rbrack\;.\nonumber\\
\end{eqnarray}
In a similar way we determine the ground state energy for the normal phase
($\Delta=0$). There are only two stable solutions: with $b^2 = 2|\delta|$,
corresponding to the state of strongly correlated electrons, in which only
single occupancy is allowed, and with $b^2= 1$, corresponding to the
diamagnetic metal, in which single occupancy is forbidden. There is not
present the weakly correlated phase (with $2|\delta|<b^2<1$), in which occurs
sites with two electrons as well as sites with one electron. The free energies
for these two stable situations are
\begin{eqnarray}\label{20}
F_1^N/N =&& -\frac{16t|\delta|}{\pi(1+2|\delta|)}\sin\frac{\pi}{2}(1-2|\delta|)
+U (|\delta|+\delta)\nonumber\\ 
&&\qquad\qquad\qquad\qquad \mbox{ for } U>U_{12}\; ,\\
\label{21}
F_2^N/N =&& U\frac{n}{2}\qquad\qquad\qquad\mbox { for } U<U_{12}\;,
\end{eqnarray}
where
\begin{equation}\label{22}
U_{12} = -\frac{32t|\delta|}{\pi(1-4\delta^2)}\sin\frac{\pi}{2}(1-2|
\delta|)\;.
\end{equation}
Putting $\Delta=0$ ($|\overline\delta|=|\delta|$) in Eq. (\ref{19}) we get
the critical value of $U$, below which the superconducting phase is stable.
The phase transition is the second order to the normal phase of strongly
correlated electrons. The excitation spectrum in the superconducting phase has
no gap, however, there are stable Cooper pairs ($\Delta\neq 0$) as it is seen
from the energy difference $F^N-F^S$ shown in Fig.\ \ref{f1}. For $U<-4t$ the
free energy of the superconducting state and the normal state are the same.
This degeneration may be removed by any interaction stabilizing one of the
phases. We will show later that a deviation from $X=t$ stabilizes
superconductivity.

	The solutions in the Hartree-Fock approximation may be also presented
in the parametric form with the parameter s. The superconducting state has the
free energy at $T=0$
\begin{eqnarray}\label{23a}
F_{HF}^S/N =&& -\frac{8ts}{\pi}\sin \frac{\pi}{2}(1-2s)+Us^2+\frac{U}{4}(1+4
\delta),\\
\label{24a}
U=&& 4t[\frac{1}{\pi s}\sin \frac{\pi}{2}(1-2s)-\cos \frac{\pi}{2}(1-2s)]\;.
\end{eqnarray}
and the order parameters are given by
\begin{eqnarray}\label{23b}
\Delta_0=&& <c_{i\downarrow}c_{i\uparrow}>=\sqrt{s^2-\delta^2}\;,\\
\label{24b}
\Delta_1=&& <c_{i\downarrow}c_{j\uparrow}> =-\frac{1}{\pi}\sin \frac{\pi}{2}(1
-2s)\;.
\end{eqnarray}
This state is stable for U lower than
\begin{equation}\label{23c}
U_S^{HF}= 4t[\frac{1}{\pi |\delta|}\sin \frac{\pi}{2}(1-2|\delta|)-\cos \frac{
\pi}{2}(1-2|\delta|)]\;.
\end{equation}
The free energy of the normal state is that given by Eq.(\ref{23a}) with $s=|
\delta|$.

	For comparison the exact results~[\cite{Elsser,Aligia94}] for the ground
state energy in the 1D chain are
\begin{eqnarray}\label{23}
F_{ex}/N =&& -\frac{1}{2\pi}\sqrt{16t^2-U^2}+\frac{U}{2}[n-\frac{1}{\pi}
\arccos(-\frac{U}{4t})]\nonumber\\
&&\qquad\qquad\qquad\mbox{ for } -4t\le U\le U_S^{ex}\; ,\\
\label{24}
F_{ex}/N =&& -\frac{2t}{\pi} \sin\pi n\qquad \mbox{ for } U>U_S^{ex}\;,\\
\label{25}
F_{ex}/N =&& U \frac{n}{2}\qquad\qquad \mbox{ for } U<-4t\;,
\end{eqnarray}
where $U_S^{ex} = -4t\cos\pi n$.

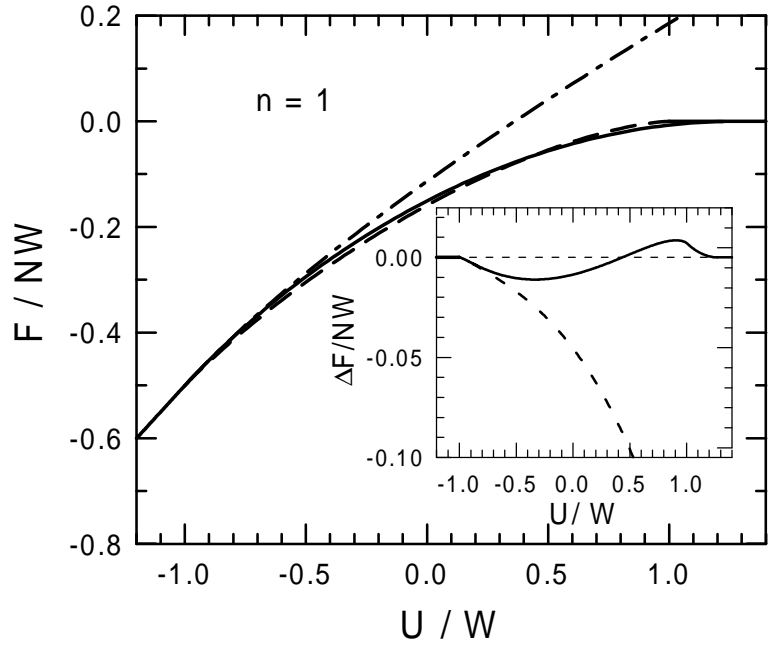
\begin{figure}
\caption{Free energy at $T=0$ for the 1D Hubbard model with correlated hopping at
$X=t$, $n=1$, calculated by means of the SBMFA (solid curve), the HFA
(long-short dashed curve) and the exact method~[\cite{Elsser}] (dashed
curve). Insert shows the differences $F_{ex}-F^{sb}$ (solid curve) and
$F_{ex}-F^{HF}$ (dashed curve). (The bandwidth $W=4t$.)}
\label{f2}
\end{figure}

	Fig.\ \ref{f2} shows the quantitative comparison of the free energies
determined by these methods for $n=1$. The SBMFA solution (the solid curve) is
very close to the exact one (the dashed curve).  Although the SBMFA and the
HFA solutions [Eqs.(\ref{17})-(\ref{22}) and Eqs.(\ref{23a})-(\ref{23c}),
respectively] are similar in their structure, the values of the free energies
are different. The free energy for the HFA (the long-short curve in Fig.\
\ref{2}) is always higher than the exact and the SBMFA ones. The reason is in
a reduction of the hopping integral and the width of the electronic band (the
narrowing band effect) in the SBMFA. The physics in the HFA is also different:
in the normal phase may occur doubly occupied sites even for very large $U$.
For $U<-4t$ the system is highly degenerate, and all solutions [Eqs.(\ref{18}),
(\ref{21}), (\ref{23}) and (\ref{30})] (as well as the free energy for CDW at
$n=1$) are identical $F/N=Un/2$. It is energy of localized electron pairs, in
an absence of hopping process, and therefore, it corresponds to an insulating
state. The Drude weight determined~[\cite{Arrachea}] by changing the twisted
boundary conditions of the 1D model equivalent to (\ref{1}) (and similar to
(\ref{7})) is $D_c=0$ for $U<-4t$ and any $n$, indicating on the insulating
phase. Above $U_S$ the stable solution is the strongly correlated state. The
same free energy has the ferromagnetic state with saturated magnetic moment
$m=n/2$ (as well as the antiferromagnetic state in the case $n=1$). In the
intermediate regime $-4t<U<U_S$, may also occur a nonuniform ordering, in
particular, the state with two separated phases: the phase with only doubly
occupied sites and the phase with only singly occupied sites. The stability of
a such two-domain system may be analyzed by the Maxwell construction, using
the single domain results for the SBMFA, Eqs.(\ref{20}) and (\ref{21}). The
free energy is then exactly the same as for the uniform superconducting state
given by Eq.(\ref{18}). One can show that the results obtained in the exact
manner~[\cite{Elsser,Aligia94}] [Eqs.(\ref{23})-(\ref{25})] have the same
properties (see also Ref.[\onlinecite{Arrachea94a}]).

	The SBMFA may be apply also for the systems of higher dimensions. The
results depend on the density of states (DOS) and differ slightly. For
illustration we present below the ground state characteristics for the
rectangular DOS ($N(E) = 1/W$ for $|E|\le W/2$):
\begin{eqnarray}\label{26}
F^S/N =&&  -\frac{W}{8}(1-\frac{U}{W})^2+U\delta\quad\mbox{for}\;\;
 -W<U<U_S\;,\nonumber\\
F^S/N =&& U\frac{n}{2}\quad\mbox{for}\;\; U<-W\;,\\
\label{27}
F^N_1/N =&& -W|\delta|(1-2|\delta|)+U(|\delta|+\delta)\nonumber\\
&&\qquad\qquad\qquad\qquad\mbox{for}\;\; U>-2|\delta|W\;,\nonumber\\
F_2^N/N =&& U\frac{n}{2}\quad\mbox{for}\;\; U<-2|\delta|W\;,
\end{eqnarray}
where $U_S= W(1-4|\delta|)$. These results may be compared with those analyzed
above as well as with those for a d-dimensional hypercubic lattice taking the
bandwidth $W=4dt$. The HFA applied to the Hamiltonian (\ref{1}) with the
rectangular DOS gives qualitatively different results:
\begin{eqnarray}\label{28}
&&F^S_{HF}/N= \frac{U}{4}(1+4\delta)+\nonumber\\
 &&\frac{1}{108W^2}[U-\sqrt{U^2+3W^2}]
[6W^2+U(U-\sqrt{U^2+3W^2})]\nonumber\\ 
&&\qquad\qquad\qquad\qquad \mbox{for}\;\; -W<U<U_S^{HF}\;,\\
&&F^S_{HF}/N= \frac{U}{2}n\quad\quad\mbox{for} \quad U<-W\;,\\
\label{29}
&&F^N_{HF}/N= \frac{U}{4}n^2-\frac{W}{2}|\delta|(1-4\delta^2)\;,
\end{eqnarray}
and $U_S^{HF}=  W(1-12|\delta|^2)/(4|\delta|)$.

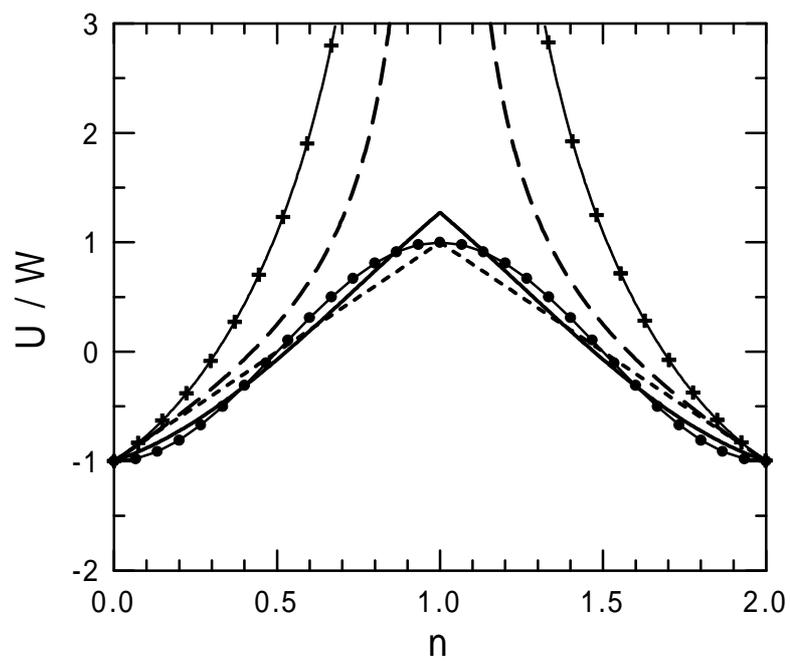
\begin{figure}
\caption{Stability diagram for the superconducting state of the Hubbard model
with correlated hopping at $X=t$ determined by the SBMFA for the rectangular
DOS (short dashed curve) and the 1D chain (solid curve), the exact method for
the 1D chain (curve with dots), and by the HFA for the rectangular DOS (long
dashed curve) and for the 1D chain (curve with crosses).}
\label{f3}
\end{figure}

	Fig.\ \ref{f3} summarizes our results. There are exhibited the
dependencies of $U_S$ vs. the electron concentration $n$ for the different
approaches. The HFA curves (the long dashed curve and the curve with crosses)
are completely different than the others. Of course, in the HFA there is no
place for the strongly correlated normal state with only singly occupied
sites. With an increasing deviation from the point $n=1$ (increasing $|n-1|$)
all the curves in Fig.\ \ref{f3} become closer to each other, correlations
between electrons become weaker.

\subsection{The Mott Transition in the Hubbard Model with Correlated Hopping:
The $n=1$ Case}

	The case of the half-filled band ($n=1$) we would like to treat
separately as it is a discontinuity point of the model. Moreover, for $X=0$
(in the Hubbard model without correlated hopping) there is the Mott transition
as a function of $U$ from a metallic phase to an insulating phase, in which
double site occupancy is forbidden. We have already learnt (in the previous
section) that at $X=t$ the strongly correlated state may occurs for any $n$
and it is metallic apart from the case $n=1$ corresponding to an insulator.
The Mott transition is then from the superconductor to the insulator. The
exact diagonalization calculations for the 1D chain performed by Arrachea et
al.~[\cite{Arrachea94a}] at $X=-t$ indicate also such a transition.

	Our procedure is similar as in the previous section. We perform our
further calculations for the rectangular DOS, which are much simpler and give
qualitatively similar results as for DOS of d-dimensional hypercubic lattices.
After the transformation (\ref{16}) the free energy at $T=0$ is
\begin{equation}\label{30}
F^S/N = \frac{\rho}{2d} f^S+U\frac{b^2}{2} - 4\rho (\overline\lambda_0
\overline\delta +\overline\lambda_S\overline\Delta) \;,
\end{equation}
where
\begin{eqnarray}\label{31}
f^S =&& d(R_+ +R_-)+ \overline\lambda_0 (R_+-R_-) \nonumber\\
&&+ \overline\lambda_S^2\ln
\frac{ R_+ + \overline \lambda_0+d}{R_- + \overline \lambda_0-d}\;,\\
\label{32}
R_{\pm} =&& \sqrt{ (\overline \lambda_0\pm d)^2+\overline \lambda_S^2}\;.
\end{eqnarray}
The bandwidth of noninteracting electrons is $W=4dt$.

	First, we analyze the normal phase ($\Delta = 0$). There are possible
three type of solutions, in which: (i) double occupancy is forbidden ($b^2=0$)
for large $U$ - the Mott insulator, (ii) there are sites with two electrons as
well sites with one electron (and, of course, there are also empty sites) -
in an intermediate coupling regime , (iii) single occupancy is forbidden
($b^2=1$) - in a strong attractive regime. We find that the minimum of the
free energy (\ref{30})-(\ref{32}) in the normal phase is:
\begin{eqnarray}\label{33}
F^N/N =&& 0\qquad\qquad\qquad \mbox{for}\quad U>U_{Mott}\;,\\
\label{34}
F^N/N =&&  -d|t-X|(\frac{U}{8d|t-X|}-1)^2\nonumber\\
&&\qquad\qquad\qquad\qquad \mbox{for}\quad |U|<U_{Mott}\;,\\
\label{35}
F^N/N =&& \frac{U}{2} \qquad\qquad\qquad \mbox{for} \quad U<-U_{Mott}\;,
\end{eqnarray}
where the critical  value for the Mott transition is $U_{Mott}= 8d|t-X|$. It
is seen that the results are the same as in the absence of the correlated
hopping ($X=0$) if we use a renormalized hopping integral $|t-X|$.  The
average concentration of doubly occupied sites is $b^2=\frac{1}{2}(1-\frac{U}
{8d|t-X|})$ and monotonicly increases from $b^2=0$ at $U=U_{Mott}$ and reaches
$b^2=1$ at $U=-U_{Mott}$.

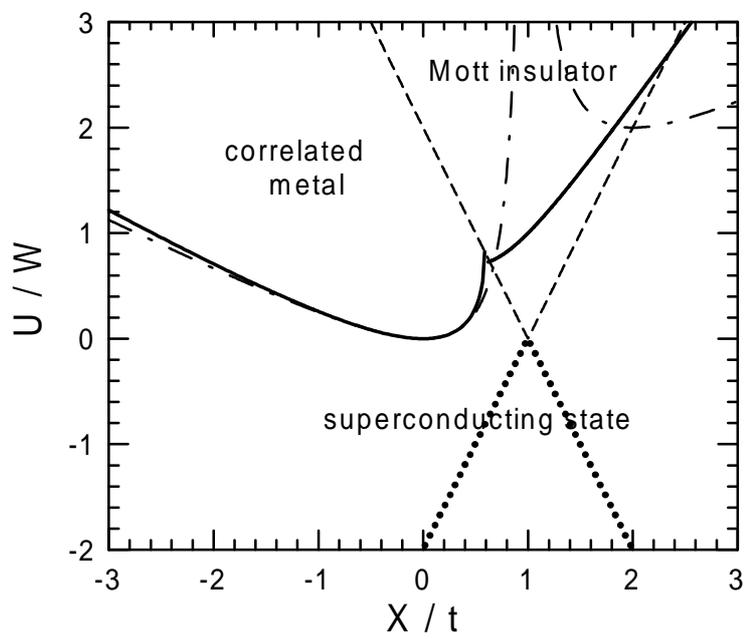
\begin{figure}
\caption{Stability diagram for the superconducting phase of the Hubbard model
with correlated hopping at $n=1$ determined for the rectangular DOS by the
SBMFA (below the solid curve). The solutions for the normal phase are: the Mott
insulator (above the dashed curve), the diamagnetic metal (below the dotted
curve) and the correlated metal is between them. For comparison, the
superconducting solutions in the HFA are stable below the long-short dashed
curve.}
\label{f4}
\end{figure}

	The analyze of the superconducting phase is now more complicated as we
have to find the minimum of $F$, Eq.(\ref{30})-(\ref{32}), which is a function
of the parameters $b^2$, $\Delta$ for a given $U$ and $n=1+2\delta$. If we
assume that the transition to the superconducting state is the second type we
can expand $F$ in a series of $\Delta$ and find $U_S$ - the critical value of
the parameter $U$ below which superconductivity is stable . There are two
solutions corresponding to the transition from the metallic phase:
\begin{equation}\label{36}
U_S = 8d(|t-X|-\sqrt{t^2-2tX+X^2/2})
\end{equation}
and from the insulator
\begin{equation}\label{37}
U_S = 4d\sqrt{t^2-2tX+2X^2}\;.
\end{equation}
The stability diagram for the superconducting phase is presented in Fig.
\ref{f4}, where the solution (\ref{36}) and (\ref{37}) are shown by the solid
curves. The dashed line shows the stability regions in the normal phase.  In
the region of small $X$ the phase transition to the superconducting state is
from the metallic state; for larger values of $X$ the transition is direct
from the Mott insulator. For comparison, the critical value $U_S^{HF}=2dX^2/
|t-X|$ obtained in the HFA is shown in Fig.\ref{f4} by the long-short dashed
curve (see Appendix C for details). The significant differences between the
both approaches are for $X>0.5t$.

	Our studies do not confirm the prediction by Arrachea et al.~
[\cite{Arrachea94a}] the superconductor-to-Mott insulator transition  at $X=-t$.
They performed exact diagonalization calculations  for the 1D chain, in which
the Mott insulator is the stable solution for any positive $U$. An comparison
with exact results obtained~[\cite{Marsiglio96}] by the Bethe Ansatz for the
attractive Hubbard model (at $X=0$) shows that some fluctuations specific for
the 1D system are not taken into account in the SBMFA.~[\cite{Bak}] The
superconductor-to-Mott insulator transition in the 1D chain for $X<0$,
obtained by Arrachea et al.~[\cite{Arrachea94a}], may not occur in systems of
higher dimension, where one expects the Mott transition for a finite value of
$U$ and the SBMFA works better.

	The stable solutions and their ground state characteristics are, in
general, determined numerically. In Fig.\ref{f5}a we present the free energy
difference $\Delta F = F^N - F^S$ between the normal and the superconducting
state vs. the parameter $U$ for a few given values of $X$. The curves for
$X=0.8t$, $1.2t$ and $1.5t$ correspond to the solutions of the superconducting
state in the strongly correlated regime, where double occupancy is forbidden.
The $\Delta F$ has a maximum at $X=0$ for all these cases. We see that at
$X=t$ the free energy for the superconducting and the normal phase is the
same for large attraction ($U<-W$), but any deviation from $X=t$ favors the
superconducting solution ($\Delta F > 0$ for any $U<U_S$). At $X=0.5t$ and
$X=-0.8t$ the system is in the weakly correlated regime, where the
metal-to-superconductor transition occurs. The maximum of $\Delta F$ is now
at ca. $2|1-X/t|W$ (the dotted curve has its maximum outside the figure at
$U=-3.1W$), similarly as in the pure Hubbard model ($X=0$) (see
Ref.[\onlinecite{rob}] for details). The $U$ dependence of the gap in the
excitation spectrum, $E_g = 4\rho |\overline\lambda_S|$, is also different in
the both regions. It is visualized in Fig.\ref{f5}b. For the strongly
correlated case (see Appendix A for an analytical solution) the effective
bandwidth $W_{eff}= 4d\rho$ as well as the gap in the excitation spectrum $E_g$
linearly increases with $U_S-U$ (see Fig.\ref{f5}c and \ref{f5}b). Therefore,
the value of $E_g$ is relative small in a wide region of $U$. In the weakly
correlated region and for $U$ close to $U_S$ the energy gap $E_g$ has an
exponential dependence. It is similar to the Hartree-Fock solution, but the
value is reduced by the factor $\gamma= 0.47$.~[\cite{rob}]

	In Fig.\ref{f6}a the difference of the free energies of the normal and
the superconducting phase is shown as a function of $X$ for different values
of $U$. The relative stability of the superconducting phase is a nonmonotonic
function and its character may be quite different for the different values of
$U$. The dependence of the gap $E_g$ in the excitation spectrum is shown in
Fig.\ref{f6}b. $E_g$ has the maximum in the weakly correlated region, reaches
zero at $X=t$ and monotonicly increases for larger $X$. The small jump around
$X\approx 0.65t$ is at the crossover point from the region of weakly
correlated electrons to the region of strongly correlated electrons. The
dependence of $W_{eff}$ on $X$ is shown in Fig.\ref{f6}c.

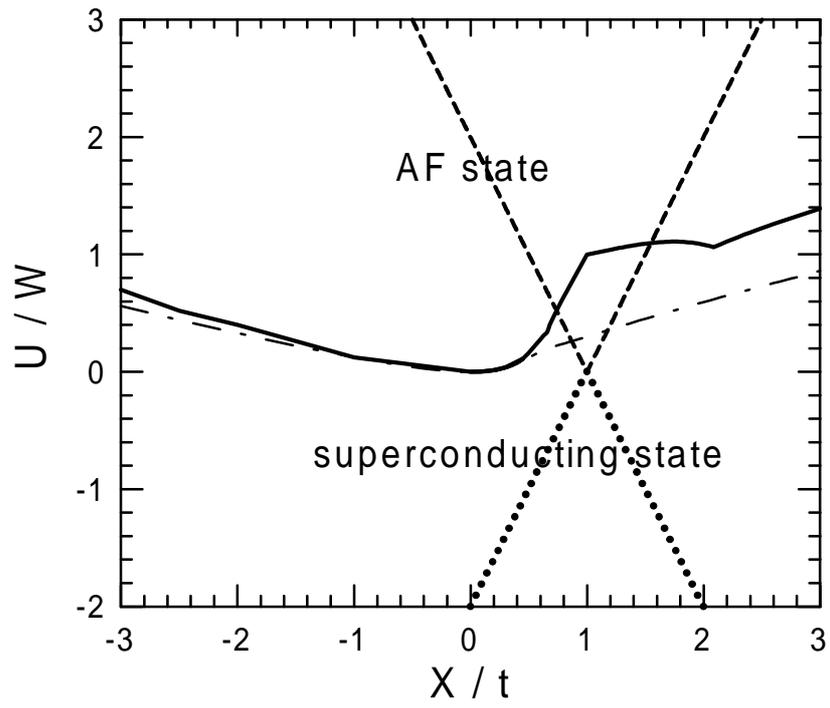
\begin{figure}
\caption{Stability diagram for the superconducting and the antiferromagnetic
phase of the Hubbard model with correlated hopping at $n=1$ determined for the
rectangular DOS by the SBMFA (solid curve). The dashed curve separates the
normal phases: the Mott insulator (above) and the correlated metal (below);
the solution for the diamagnetic metal is below the dotted curve. For
comparison, the boundary line between the AF and superconducting phase
obtained in the HFA is shown by the long-short dashed curve.}
\label{f7}
\end{figure}

	Magnetic properties of the Hubbard model with correlated hopping are
apart from our main interest in the present paper, but at $n=1$ the magnetic
ordering may significantly influence superconductivity, especially for a
positive value of $U$. Here, we want to present our studies of the relative
stability of the antiferromagnetic (AF) and the superconducting state in the
considered model for $n=1$. To analyze the stability of the AF and the
superconducting state is needed to compare their free energies at a given
point of the parameters of the model. The procedure of calculations of the
free energy of the AF state is sketched in the Appendix D. It is simpler to
determine than for the superconducting case, because one determines the free
energy only for $X=0$ and the proper rescaling of the bandwidth gives the
values for any $X$.  Fig.\ref{f7} shows the phase diagram. At $X=t$ the
transition between both phases is at $U/W=1$. The local magnetization is
then maximal $m=1/2$. For small $X$ ($X<0.66t$) as well as for large $X$
($X>2.08t$) the boundary line separates the AF phase from weakly correlated
superconducting state, whereas in the intermediate range - from the strongly
correlated superconducting state.  We also analyzed the relative stability of
the both phases in the HFA. Their free energies are always higher than those
obtained by the SBMFA. The HFA boundary is shown by the long-short dashed
curve in Fig.\ref{f7}. It is seen that for $X>0$ it lies much lower than the SBMFA line.
One can say that correlations (included in the SBMFA) more prefer
superconductivity than magnetic ordering. The CDW  phase does not occur in
Fig.\ref{f7} as its free energy is higher than that one for the AF and the
superconducting state in the whole range of the parameters (apart from the
case $X=t$ and $U<-W$ as well as $X=0$ and $U<0$, where the ground state is
degenerate).

\subsection{The case of arbitrary $X$ and $n$}

	Let us now analyze the ground state properties of the model apart from
the special points ($X=t$ and $n=1$) discussed above. Although the present
case is more complicated, we may follow in the way described already in the
Section 3a and 3b. The effective hopping integrals of the Hamiltonian
(\ref{13}) are
\begin{eqnarray}\label{38}
\tilde t =&& -\frac{(1-b^2)[(X-t)(b^2+\sqrt{b^4-4
\Delta^2-4\delta^2})+2X\delta]}{1-4\Delta^2-4\delta^2}\;,\nonumber\\
r=&& \frac{1-b^2}{1-4\Delta^2-4\delta^2}2\Delta X\;.
\end{eqnarray}
It is also needed to transform the variables $\overline{\delta}$,
$\overline{\Delta}$, $\overline{\lambda}_0$ and $\overline{\lambda}_S$
according with the formula (\ref{16}). We perform the calculations for the
rectangular DOS, for which the free energy at $T=0$ is expressed as
\begin{equation}\label{30a}
F^S/N = \frac{\rho}{2d} f^S+U\frac{b^2+2\delta}{2} -
\frac{4\rho}{d} (\overline\lambda_0\overline\delta +\overline\lambda_S
\overline\Delta) \;,
\end{equation}
where
\begin{eqnarray}\label{31a}
f^S = &&d(R_+ +R_-)+ \overline\lambda_0 (R_+-R_-) \nonumber\\
&&\qquad+ \overline\lambda_S^2\ln
\frac{ R_+ + \overline \lambda_0+d}{R_- + \overline \lambda_0-d}\;,\\
R_{\pm} =&&\sqrt{ (\overline \lambda_0\pm d)^2+\overline \lambda_S^2}\;.
\end{eqnarray}

	We study first the stability of the normal phase ($\Delta=0$). Our
task is to find, for given values of the parameters $X$, $U$ and the electron
concentration $n=1+2\delta$, the minimum of the free energy
\begin{eqnarray}\label{39}
F^N/N =&& -2d(1-b^2)|(X-t)(b^2+\sqrt{b^4-4\delta^2})+2X\delta|\nonumber\\
&&+\frac{U}{2}(b^2+2\delta)\;,
\end{eqnarray}
where the variational parameters $b^2$ is from the interval $[ 2|\delta|,1 ]$.
By comparison of the free energies we find the stability diagram for the
normal metallic state. The situation for electrons ($n<1$) and for holes
($n>1$) is presented in Fig.\ref{f8}a and Fig.\ref{f8}b, respectively. The
stable solution with $b^2=2|\delta|$ corresponds to the state of strongly
correlated (SC) electrons (holes) and it occurs (above the solid curve) for
$X>t$ when $n<1$ or for $0.5t<X<t$ when $n>1$. The lowest solid curves in the
figures correspond to the solutions obtained for $n\to 0$ and $n\to 2$,
respectively. It is seen that the boundary line for $n=0.99$ and  $n=1.01$
(the upper curves) are different than that one for $n=1$ shown by the dashed
line in Fig.\ref{f4}. The case $n=1$ is a discontinuity point, and has been
already considered. The stable solutions for the diamagnetic state with
$b^2=1$ (DM) are in the region below the dotted curve. The weakly correlated
metal is stable in the region between the solid and dotted curves.

	The normal state is characterized by the parameter $b^2$. We show its
dependence on $X$ in Fig.\ref{f9}a and Fig.\ref{f9}b, for $n=0.5$ and $n=1.5$,
respectively. There are two types of dependencies. If $U$ is large enough an
increase of $X$  leads to a monotonic decrease of $b^2$ (see the lower curve
in Fig.\ref{f9}a), which reaches the value $2|\delta|$ at the boundary with
the region of strongly correlated electrons. With a further increase of $X$
the value of $b^2$ is constant up to the transition to the region of weakly
correlated electrons, where $b^2>2|\delta|$. There is a first-order
transition at the boundary line. As the upper curve in Fig.\ref{f9}a presents,
the system may become diamagnetic ($b^2=1$), if the potential
$U$ is attractive and strong enough. One can show that the external magnetic
field aligns all spins in the strongly correlated normal phase. The HFA
calculations for the ferromagnetic state (Appendix C) give saturated
ferromagnetism with the same value of the free energy as for the strongly
correlated state. The HFA stability diagram for the ferromagnetic state
[Eq.(\ref{h9})] is different than that one in Fig.\ref{f8} as the free
energies of the normal metallic phase determined by the both method are
different. For holes (Fig.\ref{f9}b) the situation is identical under the
transformation $t\to -t+2X$ (the curves in Fig.\ref{f9}b have the same
character as those in Fig.\ref{f9}a if $X$ decreases).

	Correlated hopping changes the width of electronic band and, in
consequence, an effective mass of electrons. Fig.\ref{f9}c and \ref{f9}d
show the effective bandwidth $W_{eff}$ vs. $X$ for electrons ($n=0.5$) and
holes ($n=1.5$).  $W_{eff}$ is strongly dependent on $X$ in the weakly
correlated phase. Around the point $X=t$, $W_{eff}$ is constant and equal
to $W/(1+|n-1|)$ in the strongly correlated normal state or to 0 in
the diamagnetic state, respectively. The HFA effective bandwidth is simply
given by $W_{eff}^{HF}=|1-Xn/t|W$.

	Stability of the superconducting phase is determined numerically by
comparison of the free energies for the state of strongly correlated electrons
($b^2= \sqrt{\delta^2+\Delta^2}$) and weakly correlated electrons (the
variational parameter $b^2>\sqrt{\delta^2+\Delta^2}$). The phase transition from the superconducting to the normal state is a second-order. The stability diagram
for the superconducting state is obtained from solutions in the limit
$\Delta\to 0$. It is presented by solid curves in Fig.\ref{f10}a and
\ref{f10}b for different values of $n<1$ and $n>1$, . The dashed curves in
Fig.\ref{f10} denote the regions of the normal phase of strongly correlated
electrons (the same as in Fig.\ref{f8}). The curves corresponding to $n=0$ and
$n=2$ are the exact solutions obtained for the case of two electrons and two
holes in the Hubbard model with correlated hopping (see Appendix B for
details). The SBMFA solutions converges to these ones if $n\to 0$ and
$n\to 2$. The point $X=t/2$ is singular for the solutions with holes. For the
case of two holes the Cooper pair is stable for any value of $U$. In the SBMFA
one gets at this point the superconducting phase stable for any finite
concentration of holes ($n>1$) and also for any $U$.

	Comparison, the critical values $U_S$ obtained by means of the SBMFA
and the HFA [given by Eq.(\ref{h4}] is shown in Fig.\ref{f11}a and \ref{f11}b
for different concentration of electrons and holes, respectively. The SBMFA
and the HFA results are closer to each other for $n\to 0$ and $n\to 2$, as
correlations become weaker and weaker. Small differences are also seen in the
region $X<0.5t$. Strong correlations considerably modify the stability diagram
for $X>0.5t$, the results for the SBMFA and the HFA are qualitatively
different there.

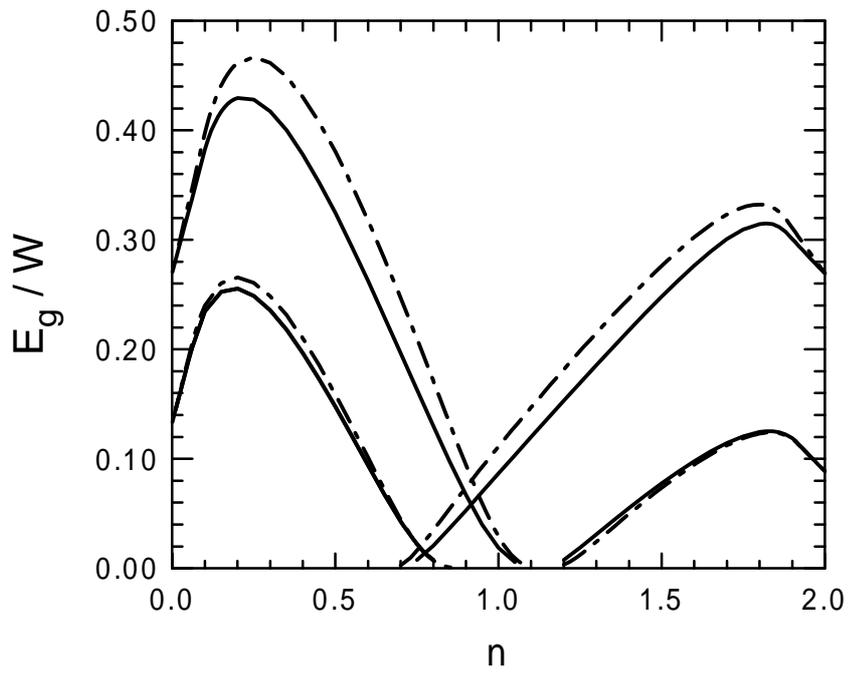
\begin{figure}
\caption{Dependence of the gap $E_g$ in excitation spectrum on the electron
concentration $n$ obtained in the SBMFA (solid curves) and the HFA (long
short dashed curve) for $X/t=0.4$, $U/W=-0.2$ and $X/t=0.4$, $U/W=0.2$ (curves
from top and the left hand side) and $X/t=-0.8$, $U/W=-0.2$ and $X/t=-0.8$,
$U/W=0.2$ (curves from top on the left hand side).}
\label{f12}
\end{figure}

	The SBMFA studies for $X=0$ have shown~[\cite{rob}] that the value of
the gap $E_g$ in the excitation spectrum can be much lower than the HFA result,
especially for $U$ close to $U_S$. We expect a similar reduction of $E_g$ in
the presence of correlated hopping in the weakly correlated region. Our SBMFA
calculations confirm it. Fig.\ref{f12} presents dependencies of $E_g$ vs. $n$
for $X=-0.8t$ (on the left hand side) and $X=0.4t$ (on the right hand side).
The SBMFA and the HFA results have similar character, with a maximum at small
concentration of electrons (holes).

	The situation for $X>0.5t$ is different, as it shows Fig.\ref{f13}a.
The superconducting state is stable for n close to 1, in contrast to the case
$X<0.5t$ from Fig.\ref{f12}. There are the weakly correlated and the strongly
correlated superconducting states for $n>1$ and $n<1$, respectively. The
difference of the free energies of the normal and the superconducting state,
$F^N-F^S$, exhibited in Fig.\ref{f13}b indicates, that the stability of the
both phases depends in a different way on $X$ and $U$. For large $U$ the
stable solution may be only the strongly correlated superconducting state. It
is shown by the dashed curve, which corresponds to $X=3t$ and $U=4W$. Taking
$t=0.1$ eV we get $X=0.3$ eV and $U=4.8$ eV for a simple cubic lattice. Such
the parameters are quite realistic for cuprates. The difference $F^N-F^S$ as
a measure of the condensation energy of the superconducting state is
proportional to the critical temperature $T_c$. For these parameters we estimate
the value of
$T_c$ of the order of 100K. From Fig.\ref{f7} we see that at $n=1$, $X=3t$ and
$U=4W$ the stable solution is the AF state, which should be also placed in
Fig.\ref{f13}. It resembles the phase diagram found for cuprates, where the AF
state is destroyed  and superconductivity emerges with a decreasing number of
electrons in the system.

\section{Discussion and Final Remarks}

       In this paper the role of the term of correlated hopping added to the
Hubbard Hamiltonian has been studied by means of the SBMFA for the temperature
$T=0$. We have shown that the SBMFA stability of the superconducting phase is
qualitatively similar to that determined in the exact way for $X=t$. This case
is highly degenerate as electronic correlations reduce the width of the band
to zero for $U<-W$. We found the strongly correlated phase (with only singly
occupied sites) as a stable solution also apart from the point $X=t$, namely,
for $n<1$ at $X\ge t$, and for $n>1$ at $t/2\le X \le t$ if the onsite
interaction $U$ is strong enough. This state is metallic, in which all spins
may be polarized in the presence of an infinitesimal external magnetic field.
It means saturated ferromagnetism. The SBMFA stability boundary for this state
is lower than the value of $U$ determined by the optimal ground state
approach as the sufficient conditions for stable ferromagnetic
solutions~[\cite{OGSA}].
At the half-filled band the strongly correlated state is
the Mott insulator, however, it is not the state with the lowest energy. The
ground state is either the AF state or the superconducting state, depending on
the value of $U$. The phase diagram for the superconducting phase is close to
the HFA one only for negative and small positive values of $X$ at any $n$, and
for arbitrary values of $X$ at $n\approx 0$ or $n\approx 2$. For $X<0$ the
superconducting state is stable at small concentration of electrons, whereas
for $0<X<0.5t$ at small concentration of holes.  In the region $X>0.5t$ there
are phases of the strongly correlated type. The stable solutions for the
weakly and the strongly correlated superconducting phase occur close to $n=1$.
The $U$ dependence of the gap in the excitation spectrum $E_g$ is also
different there, it exhibits a nonexponential character. For the parameters
corresponding to cuprates the estimated value of $T_c$ is of the order of 100K.
Our phase diagram, with the AF phase at $n=1$ and the superconducting phase
for hole doping ($n<1$), resembles that one obtained in cuprates.

	 The model analyzed by Hirsch and Marsiglio~[\cite{Hir89}] differs
slightly from that one considered above. They performed the particle-hole
transformation of the Hamiltonian (\ref{1}) with simultaneously done
interchange ${\bf k}=0$ and ${\bf k}={\bf Q}$ in the wavevector space
${\bf k}$, i.e. $c_{i\sigma}^{\dag}\to (-1)^i c_{i\sigma}$, after which
$t\to t-2X$ and $X\to -X$ (see also Ref.~[\onlinecite{Arrachea94a}] for the
analyze of symmetry properties of the model). In further considerations the
hopping integral was taken~[\cite{Hir89,Hir95}] as equal to $t$ and independent
on $X$. Their studies~[\cite{Hir89,Hir95}] correspond to the case $X<0$ in
the present work, where the weakly correlated superconducting phase occurs.
The SBMFA boundary curve for the stability of the superconducting state is a
little bit higher than the HFA result (see Fig.\ref{f11}). It is in agreement
with exact diagonalization calculations showing also a small
increase.~[\cite{Hir95,Parola}] The SBMFA value of the energy gap $E_g$ is
smaller than the HFA one. The same reduction is for the
difference of the free energies, and consequently, for the thermodynamical
critical magnetic field $H_c$. In the strong attraction regime we expect
superconductivity of local pairs (condensed bosons). The coherence length
$\xi_{GL}$ is then of the order of the lattice constant, and the $H_c$
decreases with increasing $U_S-U$ in contrast to the HFA, in which the
$H_c^{HF}$ always increases.

	The Hubbard I approximation~[\cite{H}] applied to the model was
performed by Das and Das~[\cite{Das}], and by Doma\'nski et al.~[\cite{Karol}].
In this approach two Hubbard subbands are separated by a gap for any value of
$U$. They~[\cite{Das,Karol}] determined the critical temperature $T_c$ for the
superconducting phase as a function of the parameters of the model and the
concentration of carriers $n$.  Although our analyze is restricted to $T=0$ we
may compare the critical value of $n$, below which, at given $U$ and $X$,
the $T_c$ becomes positive. Their results are in contrast to $n_c$ determined in
the HFA [from Eq.(\ref{h4})] as well as in the present approach (the SBMFA)
for the same parameters of the model. The reason of the discrepancy is in the
method they used, which completely neglects interactions between electrons in the
subbands even on the mean-field level.

	Our studies are restricted to the Hubbard model extended only for
correlated hopping, neglecting other n.n. terms. A most important one is the
density-density coulomb interaction, which if it is repulsive may restricts
the region of the strongly correlated normal state (see for example the exact
analyze~[\cite{Strack,Ovch}] for $X=t$) and the superconducting state.
The SBMFA correctly describes onsite electron correlations, but it is too poor
for the n.n. density-density interactions. In the representation
(\ref{2})-(\ref{4}) such the term is expressed by the operators $b$ and in the
mean-field approximation charge-charge correlations are properly taken into
account (for example the CDW state~[\cite{BB}]), whereas superconducting n.n.
correlations are completely neglected. Some improvement of the present SBMFA
is needed. We also confined our studies to the case $T=0$, where fluctuations 
of bosonic fields were neglected. One can extend the studies for $T>0$ using 
the path integral approach and taking into account the fluctuations of bosonic 
fields in a coherent potential approximation (CPA) (in an analogous way as one 
treats static~[\cite{H89}] or dynamic fluctuations for $D=\infty$~[\cite{UI}] ).

\acknowledgments
We would like to thank S. Robaszkiewicz for many illuminating discussions and
critical reading the manuscript. We also acknowledge to R. Micnas,
K. Wysoki\'nski, A.M. Ole\'s, A. Montorsi and M. Robaszewska for many useful
discussions. We are grateful for the financial support by the Polish Research
Committee of Sciences (Grant No.~2 P03B 165 10) and by the Institute for
Scientific Interchange Foundation in Torino (EC PECO Network ERBCIPDCT940027).

\appendix
\section{Superconducting solutions in strongly correlated region for
half-filled band and close to transition}

	Here, we want to present analytical solutions for the superconducting
state in strongly correlated regime for $n=1$. In this case $b^2= 2|\Delta|$.
Expanding the effective Hamiltonian parameters $\tilde{t}, r$ [Eq.(\ref{14})]
in a series of $\Delta$, we find that the effective hopping
$\rho = \Delta \rho_0$, where $\rho_0 = \sqrt{t^2-2tX+2X^2}$, and from the
condition $\partial F/\partial \Delta = 0$ the gap parameter $\Delta$ is found to be
\begin{equation}\label{a1}
\Delta = \frac{1}{4}(1-\frac{U}{U_S})\;,
\end{equation}
where $U_S= 4d\rho_0$. The linear dependence of $\Delta$ is over a wide range
of $U$ for $X$ close to $t$. It means that the effective bandwidth
$W_{eff}=4d\rho$ linearly increases with a decrease of $U$. From the
conditions $\partial F/\partial \overline\lambda_0 = 0$ and
$\partial F/\partial \overline\lambda_S = 0$ we find in the limit
$\overline\lambda_S \to 0$ the expressions for $U$
\begin{equation}\label{a2}
U=U_S(1+\frac{2\rho_0\overline\lambda_S}{d|t-X|}\log{\frac{2d}
{\overline\lambda_S}})
\end{equation}
and for the gap in the excitation spectrum
\begin{equation}
\label{a3}
E_g = 4\rho|\overline\lambda_S| = \frac{2\rho_0^2}{d^2|t-X|}
\overline\lambda_S^2\log{\frac{2d}{\overline\lambda_S}}\;.
\end{equation}
The other quantities may be expressed by means the above relations with
$\overline\lambda_S$ as the parameter.

\section{Binding energy of two electrons (holes) in the Hubbard model with
correlated hopping}

	The problem of two particles in an empty lattice may be solved exactly
for our model. It is a good test for our calculations in the SBMFA. For two
electrons on an empty lattice the trial function for formation of the Cooper
pair is (see also Ref.[\onlinecite{Marsiglio,Parola}])
\begin{equation}\label{b1}
\Psi = \sum_k f_k c_{k\uparrow}^{\dag}c_{-k\downarrow}^{\dag}|vac\rangle\;.
\end{equation}
From the Schr\"odinger equation for the Hamiltonian (\ref{1}) and the trial
function (\ref{b1}) we get
\begin{equation}\label{b2}
(E+4t\eta_k)f_k = (U+4X\eta_k)\frac{1}{N}\sum_q f_q+4X\frac{1}{N}
\sum_q f_q\eta_q\;.
\end{equation}
The eigenvalues are determined from the condition
\begin{equation}\label{b3}
1-8XI_1-UI_0+16X^2(I_1^2-I_0I_2)=0\;,\end{equation}
where
\begin{equation}\label{b4}
I_n=\frac{1}{N}\sum_k\frac{\eta_k^n}{E+4t\eta_k}\;.
\end{equation}
We change the sum over $k$ to the integral and calculate it under the
assumption of the rectangular DOS (with $|\eta_k|\le d$) the corresponding
integrals. The condition for the eigenvalue is now
\begin{equation}\label{b5}
8dt(t-X)^2=[Ut^2-EX(2t-X)]\log{\frac{E-4dt}{E+4dt}}\;.
\end{equation}
The binding energy of the Cooper pair is $E_b= E-4dt$. From this condition we
find the critical value of $U$, below which the Cooper pair is stable
\begin{equation}\label{b6}
\frac{U_S}{8dt}= -\frac{X(2t-X)}{2t^2})\;.
\end{equation}
Changing $t\to -t+2X$ we get $U_S$ for the two holes in the system
\begin{equation}\label{b7}
\frac{U_S}{8dt}= \frac{X(2t-3X)}{2t|t-2X|}\;.
\end{equation}
At $X=t/2$ the solution is singular, as the hopping for holes is equal zero.
The binding energy of the Cooper pair is then
\begin{equation}\label{b8}
E_b=\frac{1}{2}(U-\sqrt{U^2+64X^2\alpha_2})\;,
\end{equation}
where $\alpha_2= \frac{1}{N}\sum_k\eta_k^2$ and for the rectangular DOS
$\alpha_2 = d^2/3$. For the 1D chain the critical values of $U$ are the same
as those given by Eq.(\ref{b6})-(\ref{b7}) with $d=1$, although the
eigenvalues have different dependencies.

\section{The Hartree-Fock analyze of the stability of the superconducting,
antiferromagnetic and ferromagnetic state}

	The free energy at $T=0$ of the superconducting state for the model
(\ref{1}) is expressed by
\begin{equation}\label{h1}
F_S/N = -pt_{HF}+8X\Delta_0\Delta_1+U\Delta_0^2+U\frac{n^2}{4}\;,
\end{equation}
where
\begin{eqnarray}\label{h2}
t_{HF}= 2t-2Xn\;,\;\; p=\sum_{\sigma}\langle c_{i\sigma}^{\dag}c_{j\sigma}\rangle \nonumber\\
\Delta_0=<c_{i\downarrow}c_{i\uparrow}>\;,\;\;
\Delta_1= <c_{i\downarrow}c_{j\uparrow}>\;.
\end{eqnarray}
The stability conditions for this state are:
\begin{eqnarray}\label{h3}
n=&& \frac{1}{N}\sum_{k}(1-\frac{\epsilon_k}{E_k})\;,\\
p=&& \frac{1}{N}\sum_{k}\eta_k(1-\frac{\epsilon_k}{E_k})\;,\\
\Delta_0=&& -\frac{1}{N}\sum_{k}\frac{V_1+V_2\eta_k}{2E_k}\;,\\
\Delta_1=&& -\frac{1}{N}\sum_{k}\eta_k\frac{V_1+V_2\eta_k}{2E_k}\;.
\end{eqnarray}
We use the following notation: $\epsilon_k=-t_{HF}\eta_k-\overline{\mu}$,
$V_1=4X\Delta_1+U\Delta_0$, $V_2=4X\Delta_0$, $E_k=\sqrt{\epsilon_k^2+(V_1+
V_2\eta_k)^2}$ and $\overline{\mu}$ is the chemical potential. In general, the
solution of these self-consistent equations is performed numerically. The
critical value $U_S$ for the stable superconducting solution may be found
expanding the integrals in a series of small parameters $\Delta_0$ and
$\Delta_1$. The results for the rectangular DOS is
\begin{equation}\label{h4}
\frac{U_S}{W}= \frac{X}{2t^2|1-Xn/t|}[4(n-1)t+2X+2Xn-3Xn^2]\;.
\end{equation}
(see also Ref.[\onlinecite{Parola}] for the other DOS).

	The free energy of the AF state in our model is
\begin{equation}\label{h5}
F_{AF}/N= -pt_{HF}-Um^2+U\frac{n^2}{4}\;,
\end{equation}
where $m$ is the local magnetic moment. It is similar to that one in the
Hubbard model with the renormalized hopping integral $t_{HF}$. For $n=1$
and the rectangular DOS it may be expressed in the analytical form as
\begin{equation}\label{h6}
F_{AF}/N = \frac{U}{4}-\frac{W|1-X/t|}{4}\coth{(\frac{W|1-X/t|}{U})}
\end{equation}
with the magnetic moment
\begin{equation}\label{h7}
m=\frac{W|1-X/t|}{2U\sinh{(\frac{W|1-X/t|}{U})}}\;.
\end{equation}

	The free energy of the ferromagnetic ordering is
\begin{equation}\label{h8}
F_F/N= -pt_{HF}-4Xm(p_\uparrow-p_\downarrow)-Um^2+\frac{U}{4}n^2\;,
\end{equation}
where $p_{\sigma}=\langle c_{i\sigma}^{\dag}c_{j\sigma}\rangle$. For the
rectangular DOS the stable solution is for $U>U_F$, where
\begin{equation}\label{h9}
\frac{U_F}{W}= \frac{|1-Xn/t|(2-n)-2(1-n)}{n}\;.
\end{equation}
The value of the magnetic moment is then maximal  $m=n/2$ (the saturated
ferromagnetism) and the free energy is
\begin{equation}\label{h10}
F_F/N= -\frac{W}{2}n(1-n)\;.
\end{equation}
We assumed $n\le 1$. The results for $n>1$ one obtains by the proper
rescaling. The value (\ref{h10}) is, of course, the same as the free energy
for the singly occupied normal state obtained in the SBMFA.

\section{The antiferromagetism in the Hubbard model with correlated hopping
analyzed by means of the SBMFA}

	We want to find the free energy of the antiferromagnetic state in our
model, and in order to do it we need first to formulate the Hamiltonian in the
SBMFA for this state. The spin operator ${\bf S}_i$ is given by
\begin{equation}\label{af1}
{\bf S}_i=\sum_{\sigma\sigma'}\frac{1}{2}c_{i\sigma}^{\dag}
{\bf \tau}_{\sigma\sigma'}c_{i\sigma'} = \sum_{\sigma\sigma'\sigma"}
{\bf \tau}_{\sigma\sigma'}p_{i\sigma\sigma"}^{\dag}p_{i\sigma"\sigma'}\;.
\end{equation}
On a hypercubic lattice the magnetic moment is modulated with the wavevector
${\bf Q}=(\pi,\pi,...)$ and its absolute value is given by
\begin{eqnarray}\label{af2}
m=&&\frac{1}{N}\sum_{i\sigma}\sigma\langle c_{i\sigma}^{\dag}c_{i\sigma}\rangle
\exp(i{\bf Q}\cdot{\bf R}_i)\nonumber\\
=&&\frac{1}{N}\sum_{i\sigma}\sigma\langle
p_{i\sigma\sigma}^{\dag}p_{i\sigma\sigma}\rangle \exp(i{\bf Q}\cdot {\bf R}_i)\;,
\end{eqnarray}
where ${\bf R}_i$ is the lattice vector. To express the hopping term of the
Hamiltonian we use the renormalized Hubbard operators in the form (\ref{8})
neglecting off-diagonal elements of the operators $\underline p$,
$\tilde{\underline p}$, $\underline b$ and $\tilde{\underline b}$. The
fermionic part of the Hamiltonian is
\begin{eqnarray}\label{af3}
H = \sum_{k}[&&(-2\tilde t_{AF}\eta_k+\lambda_0)\sum_{\sigma}f_{k\sigma}^{\dag}
f_{k\sigma}\nonumber\\
&&+\lambda_m\sum_{\sigma}\sigma f_{k\sigma}^{\dag}f_{k+Q\sigma}]\;,
\end{eqnarray}
where the effective hopping integral
\begin{equation}\label{af4}
\tilde{t}_{AF} = (t-X)\frac{b^2[\sqrt{1-b^2+2m}+\sqrt{1-b^2-2m}]^2}{1-4m^2}\;.
\end{equation}
The free energy is given by
\begin{equation}\label{af5}
F/N = -\frac{1}{N}\sum_k E_k+U\frac{b^2}{2}-2\lambda_mm\;,
\end{equation}
where $E_k=\sqrt{4\tilde{t}_{AF}^2\eta_k^2+\lambda_m^2}$. We used the fact
that $\lambda_0=0$ for $n=1$. The stable solution is found from the minimum of
$F$, i.e. from the conditions $\partial F/\partial b = 0$, $\partial F/
\partial m = 0$ and $\partial F/\partial \lambda_m = 0$. We see that the
correlated hopping changes only the effective hopping integral
$\tilde{t}_{AF}$, and therefore, the result for any $X$ is that one for $X=0$
with the proper rescaling.

\begin{figure}
\caption{Free energy difference of the normal and superconducting state as a
function of $U$ for the rectangular DOS at $n=1$ and $X= t$ (solid curve),
$1.2t$ (short dashed curve), $1.5t$ (long short dashed curve), $0.8t$ (long
short short dashed curve), $0.5t$ (long dashed curve) and $-0.8t$ (dotted
curve) - Fig.(a). Energy gap vs. $U$ for the parameter $X= 0.5t$ (long dashed
curve) and $1.2t$ (short dashed curve) - Fig.b. It is seen the different
dependencies of $E_g$ close to the transition to metallic phase at
$U_S=0.293W$ for $X=0.5t$ and to insulator at $U_S=1.217W$ for $X=1.2t$,
respectively. Fig.c shows the change of the effective bandwidth $W_{eff}$ with
$U$ for the stable solutions at $X=0.5t$ (long dashed curve) and $X=1.2t$
(short dashed curve).}
\label{f5}
\end{figure}

\begin{figure}
\caption{Free energy difference of the normal and superconducting state as a
function of $X$ determined within the SBMFA for the rectangular DOS at $n=1$
and $U=0.3W$ (long dashed curve), $U=0W$ (solid curve), $-0.3W$ (short dashed
curve), $-0.5W$ (long short dashed curve), $-1W$ (long short short dashed
curve)- Fig.a. Energy gap $E_g$ as a function of $X$ at $n=1$ and $U=-0.3W$, $0W$,
$0.3W$ (from top) - Fig.b. In Fig.c the dependence of $W_{eff}$ vs. $X$ is
shown for the stable solutions at $U/W=-0.3$, 0.0 and 0.3 (from bottom).}
\label{f6}
\end{figure}

\begin{figure}
\caption{Stability diagram for the metallic solutions: for the strongly
correlated (SC) state (between the solid curves), the diamagnetic metal (below
the dotted curve) and the weakly correlated (WC) state (between the dotted and
solid curves) at the electron concentration $n=0$, 0.2, 0.5, 0.8, 0.99 (from
top) - Fig.a, and for $n=2$, 1.8, 1.5, 1.2, 1.01 (from bottom) - Fig.b. }
\label{f8}
\end{figure}

\begin{figure}
\caption{Dependence of the parameter $b^2$ on $X$ in the normal state for
$U= -0.75W$ and $0.25W$ (from top) at the electron concentration $n=0.5$
- Fig.a and $n=1.5$ - Fig.b. $b^2=2|\delta|=|1-n|$ means that doubly occupied
sites are forbidden, for $b^2=1$ there are only doubly occupied and empty
sites. In Fig. c and d the dependence of the effective bandwidth $W_{eff}$
vs. $X$ for $U/W=-0.75$ and $U/W=0.25$ (from bottom) at $n=0.5$ and n=1.5, respectively.}
\label{f9}
\end{figure}

\begin{figure}
\caption{Stability diagram of the superconducting state for the Hubbard model
with correlated hopping for $n=0$, 0.2, 0.5, 0.8, 0.99 (solid curve from top
for $X/t<0$ and from bottom for $X/t>0$) - Fig.a and for $n=2$, 1.8, 1.5,
1.2, 1.01 -(solid curve from top for $X<0.5t$ and from bottom for $X>0.5$)
- Fig.b. The broken curve is a boundary between the normal states of strongly
and weakly correlated electrons, below the dotted curve there are solutions
for the diamagnetic metal (as in Fig.8a and Fig.8b).}
\label{f10}
\end{figure}

\begin{figure}
\caption{Comparison of the stability diagram for the superconducting state
obtained in SBMFA (solid curves as in Fig.10) and HFA (long-short dashed
curves) for $n=0$, 0.2, 0.5, 0.8, 0.99 (solid curves - from top for $X/t<0$
and from bottom for $X/t>0$; long-short dashed curves - from top for $X/t<0$
and from bottom for $1/n>X/t>0$, two curves for $X/t>1/n$ correspond to
$n=0.99$ and 0.8) - Fig.a and for $n=2$, 1.8, 1.5, 1.2, 1.01 (from bottom for
$X/t<0$, from top for $X/t>0$ and from bottom for $X$ above the singular
point) - Fig.b. }
\label{f11}
\end{figure}

\begin{figure}
\caption{The gap $E_g$ in the excitation spectrum (Fig.a) and the free energy
difference of the normal and superconducting state (Fig.b) as a function of
$n$ determined within the SBMFA for $X=1.2t$ and $U=0.2W$ (solid curve), and
for $X=3t$ and $U=4W$ (dashed curve). The values of the dashed curve in Fig.a
are 10 times smaller.}
\label{f13}
\end{figure}

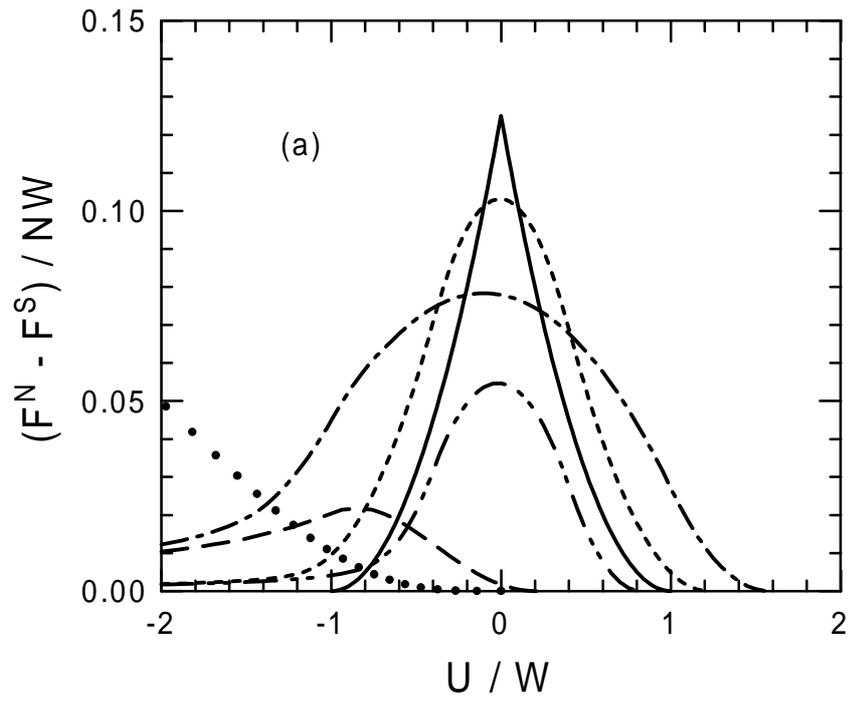
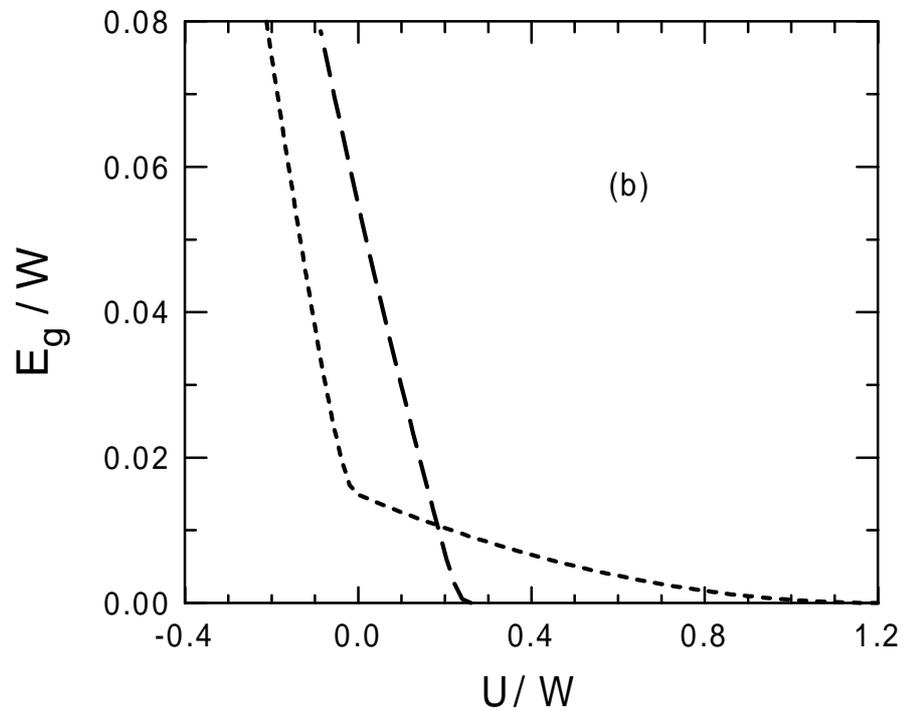
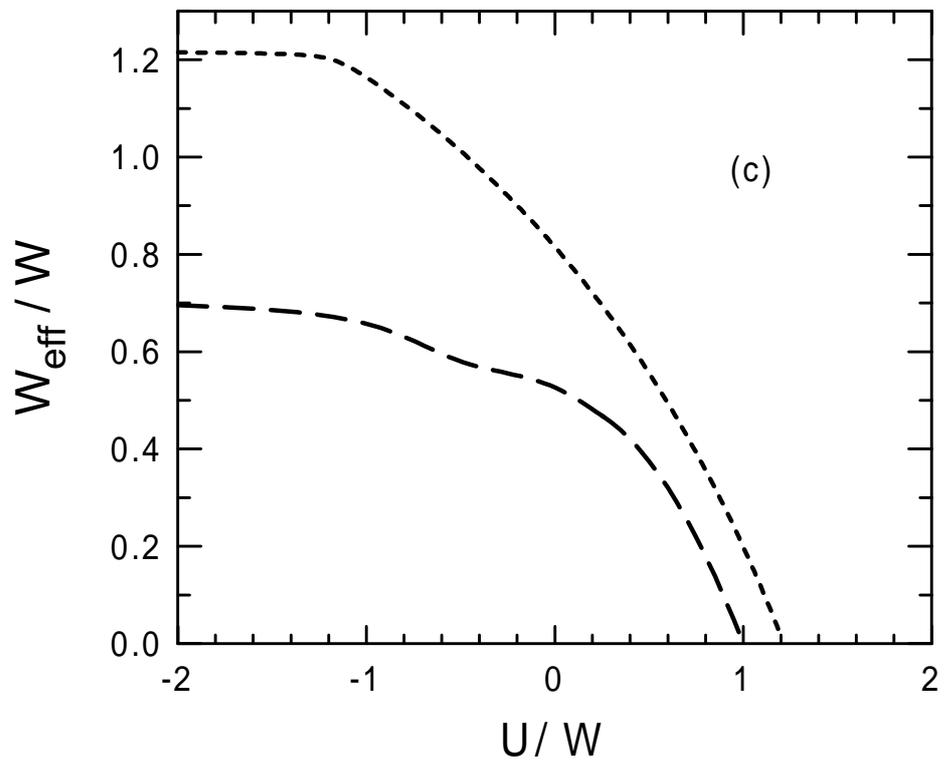
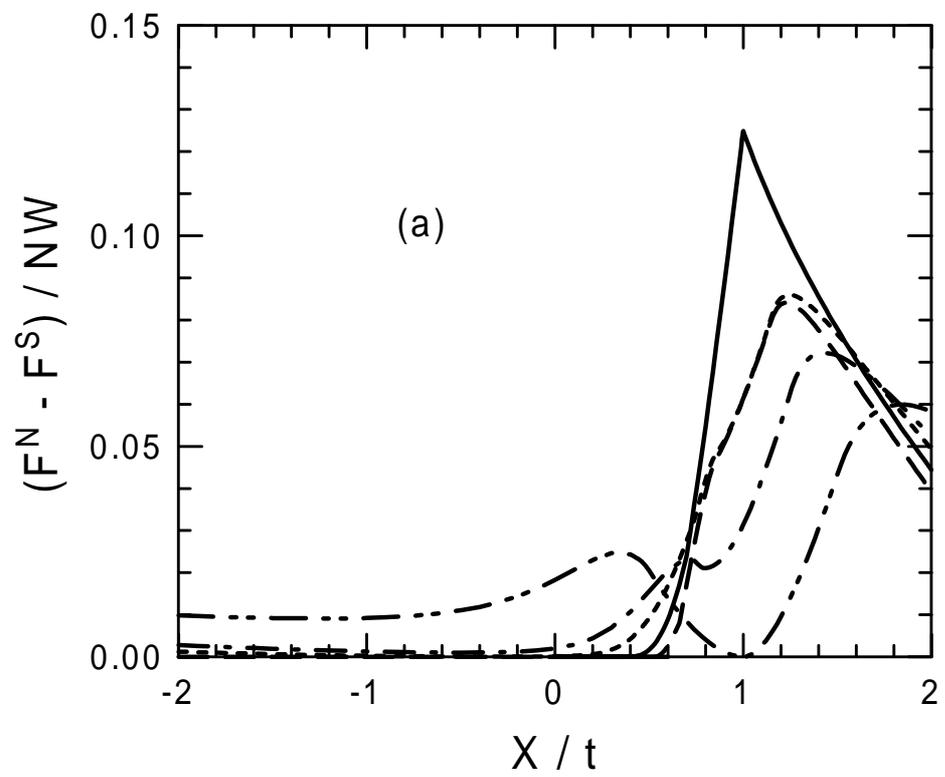
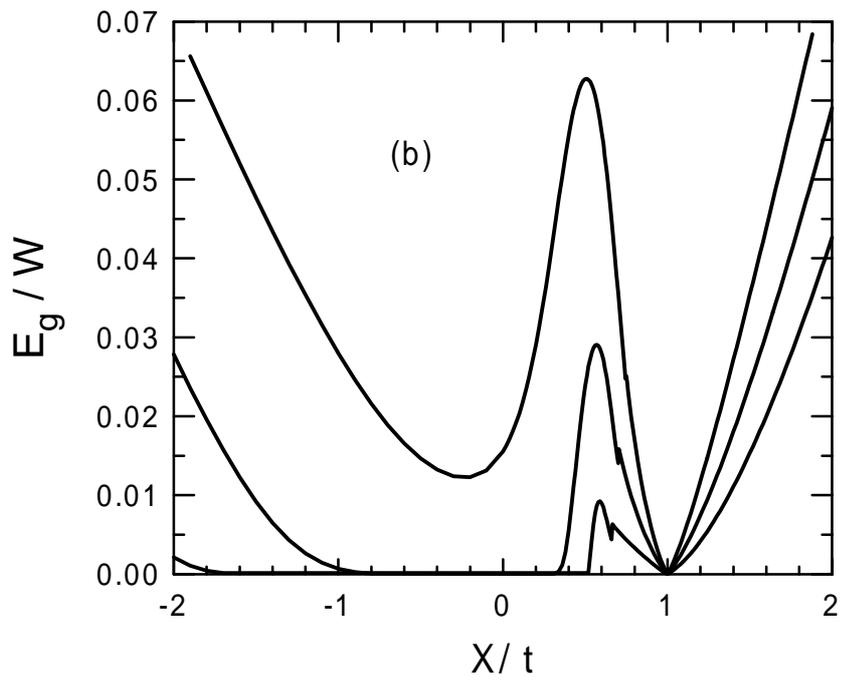
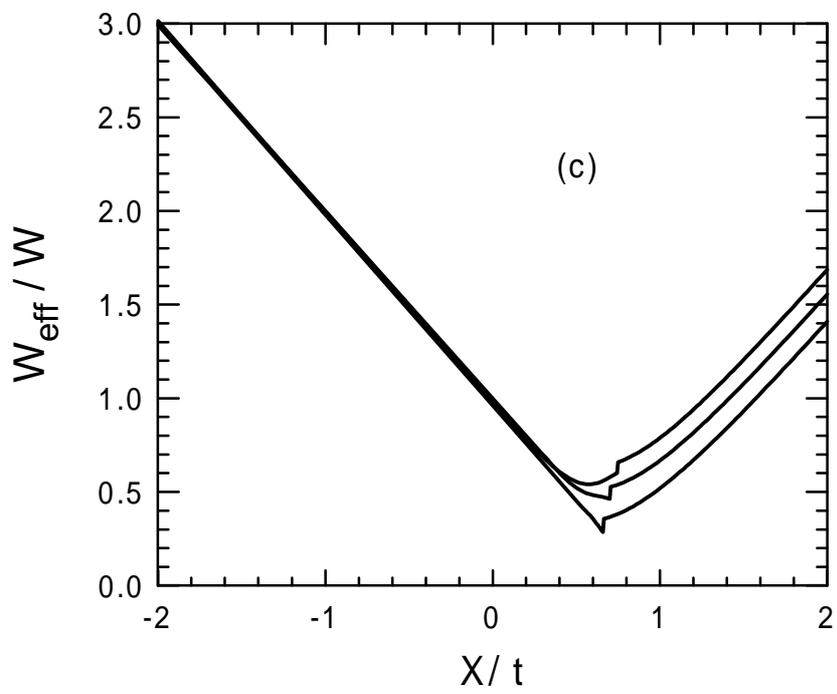
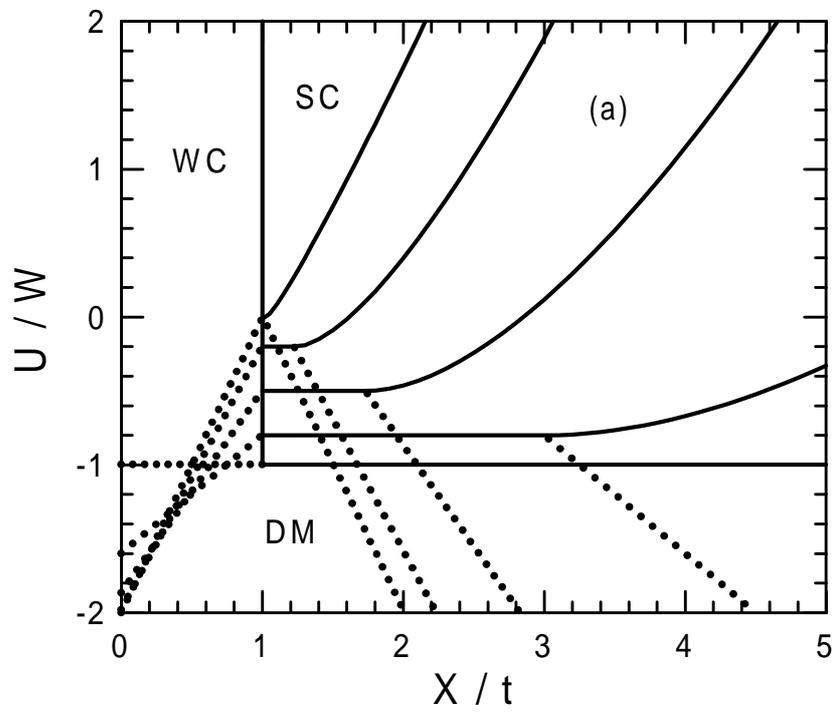
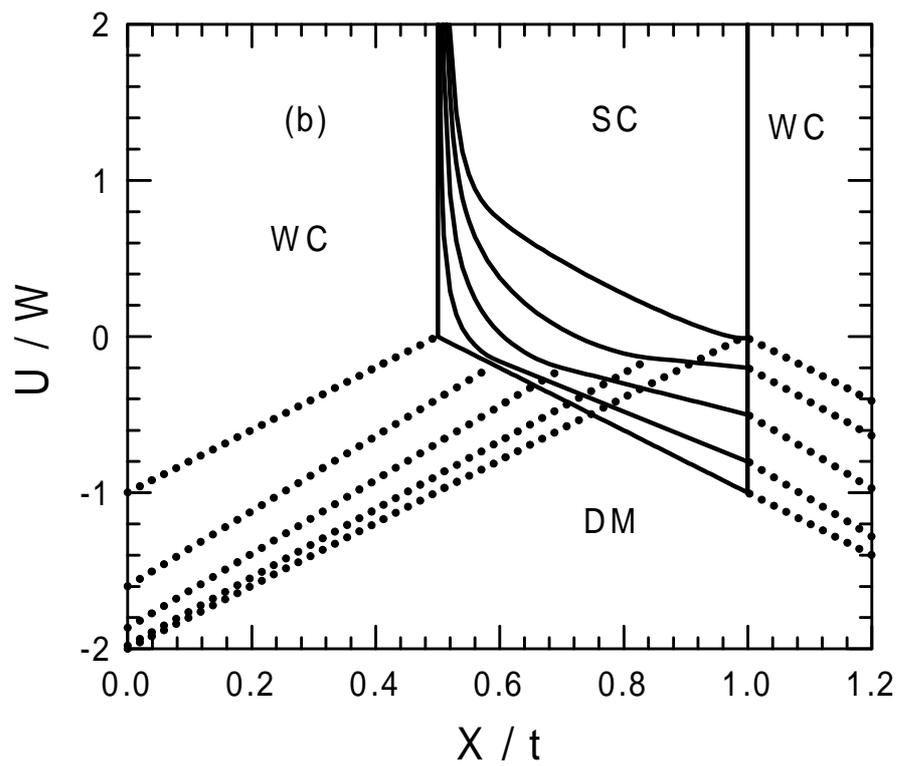
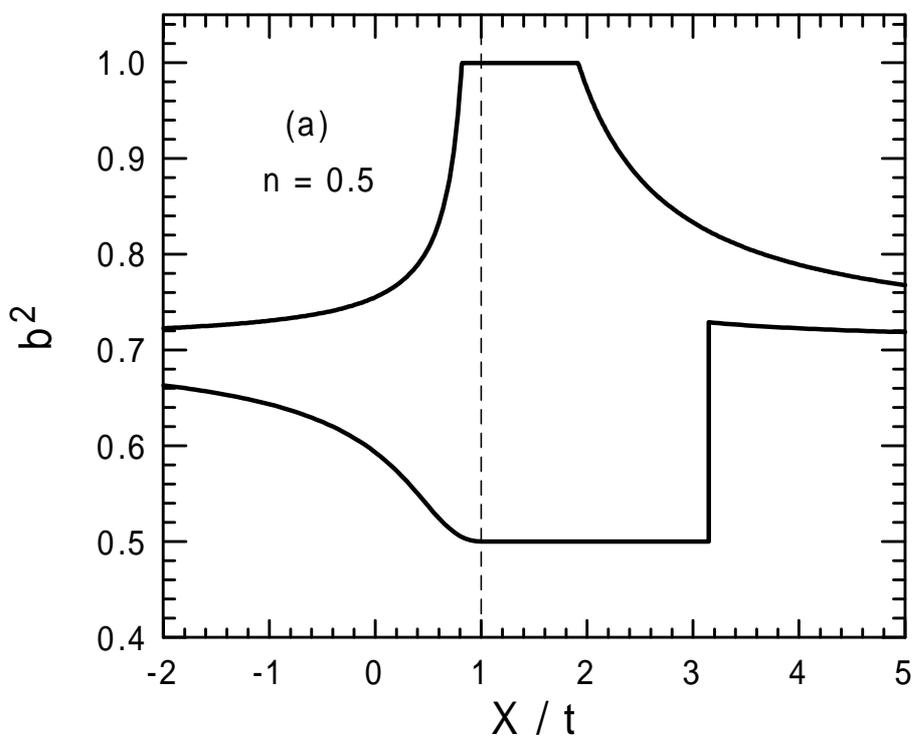
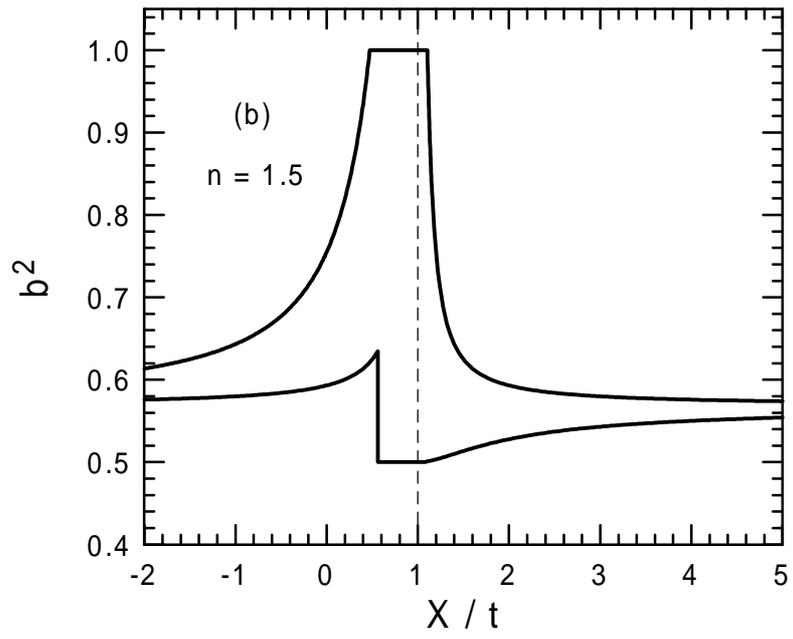
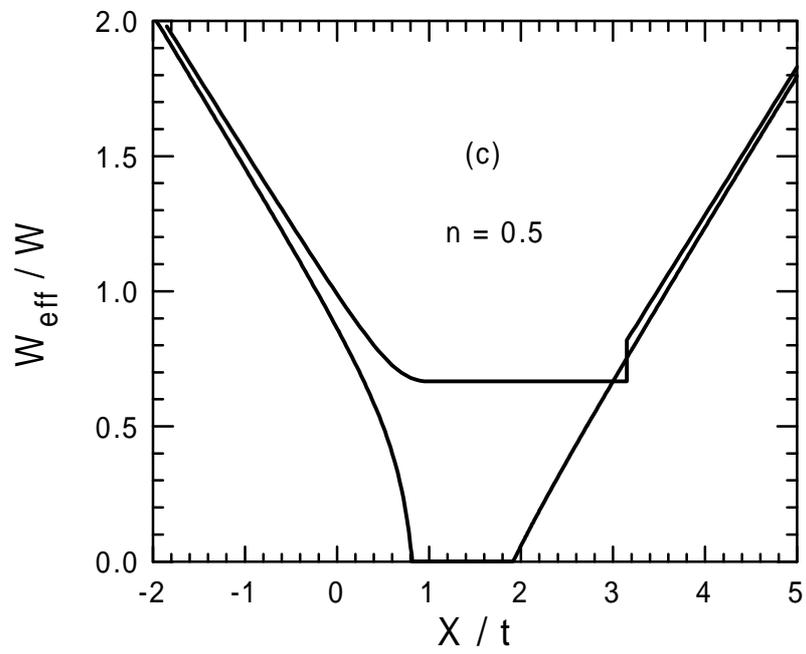
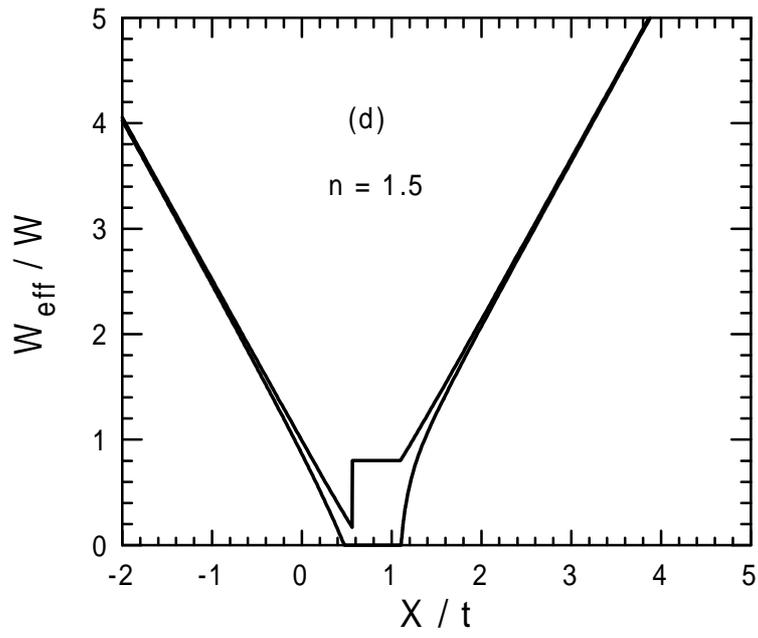
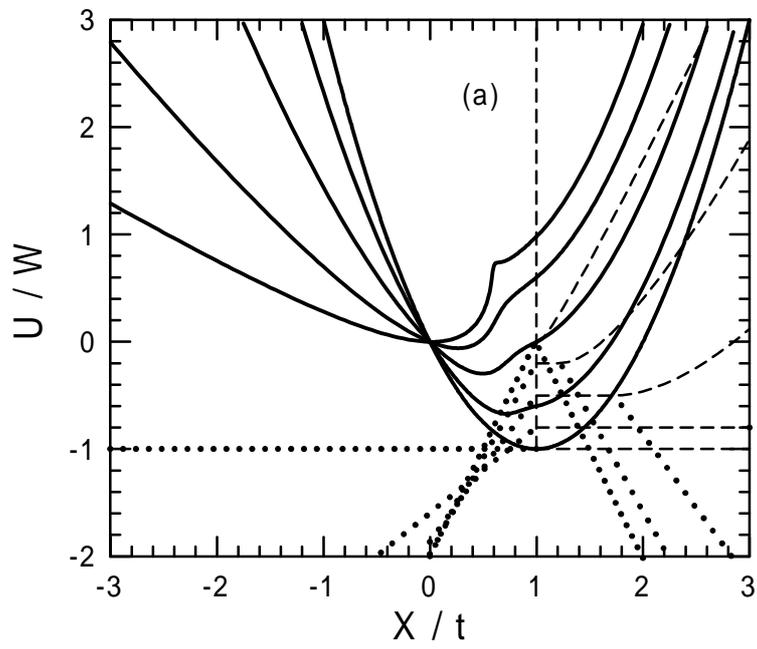
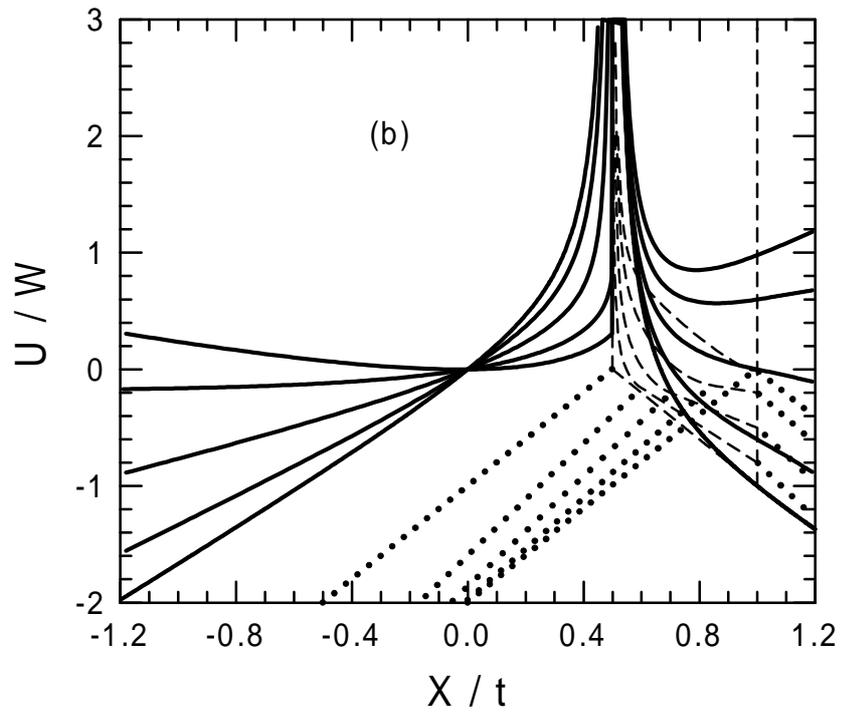
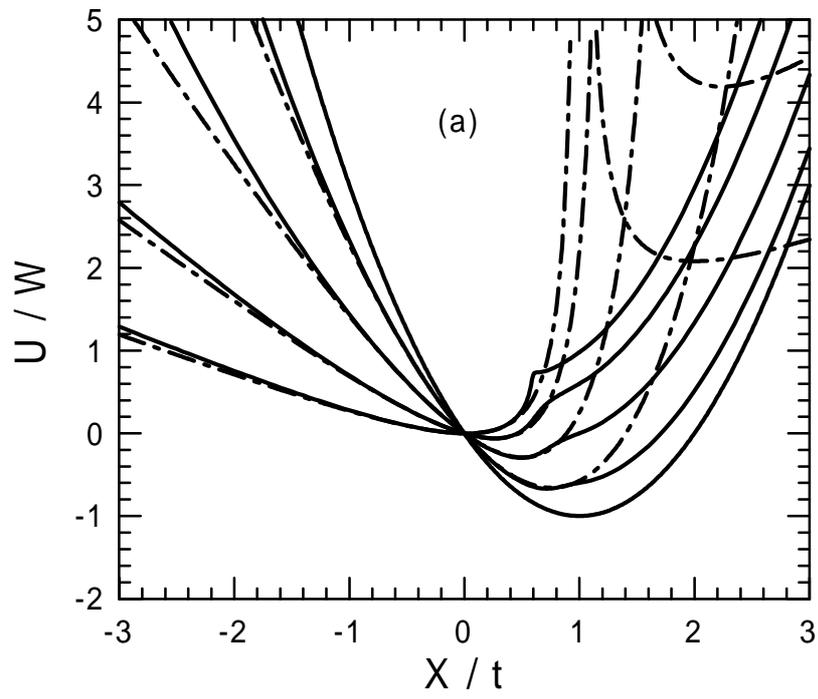
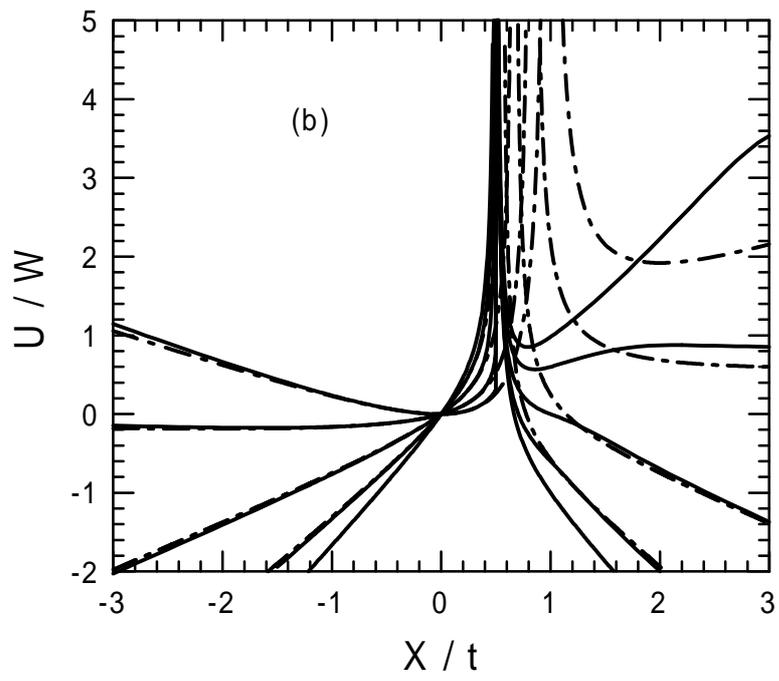
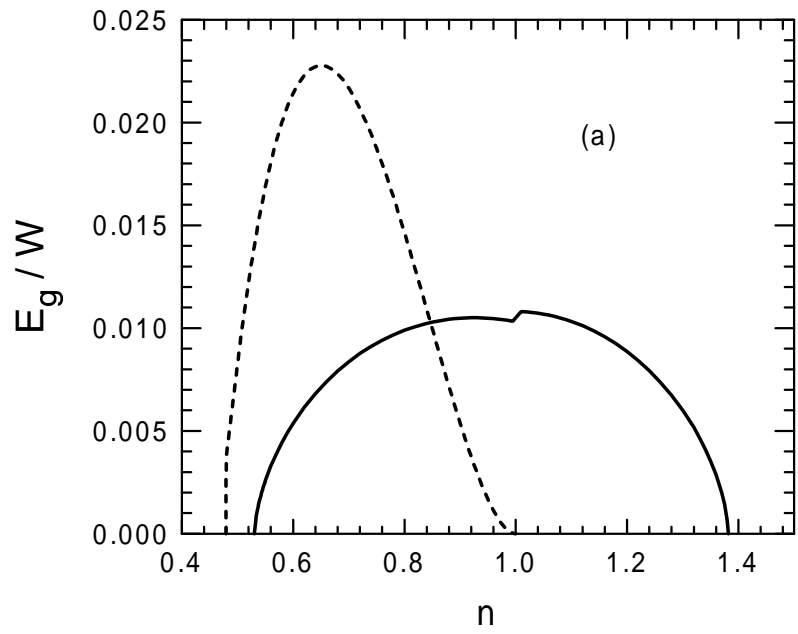
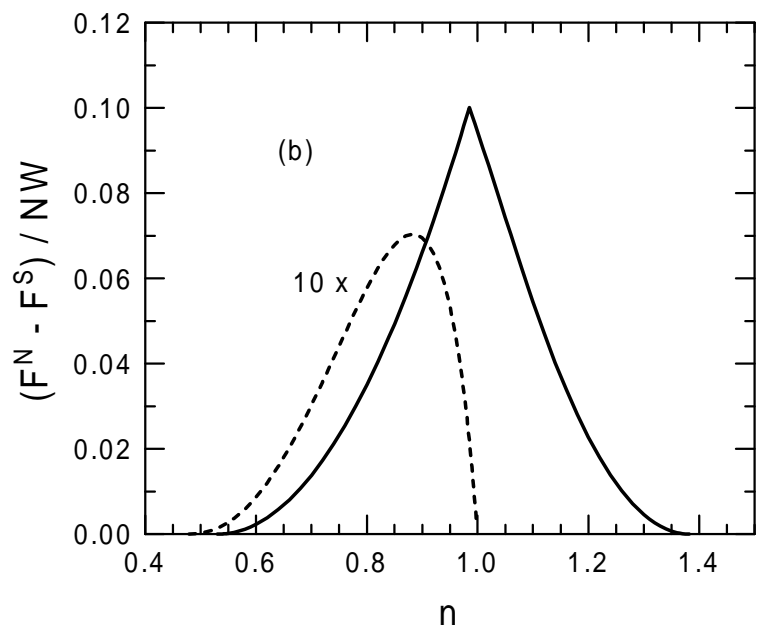
\end{document}